\documentclass[aps,floatfix,amsmath,nofootinbib,amssymb,towcolumn,superscriptaddress]{revtex4-2}
\usepackage{overpic}
\usepackage{amssymb}
\usepackage{indentfirst}
\usepackage{feynmf}   
\usepackage{slashed}  
\usepackage{cases}
\usepackage{color}
\usepackage{multirow}
\usepackage{epstopdf}
\usepackage{graphicx,color,bm}
\usepackage{epstopdf}
\usepackage{booktabs}
\usepackage[colorlinks,
citecolor=green,
anchorcolor=red,
menucolor=red,
linkcolor=red,
filecolor=red,
runcolor=red,
urlcolor=blue,
frenchlinks=red]{hyperref}

\usepackage{subfigure}
\usepackage{capt-of}
\usepackage{graphicx,amsmath}
\begin{document}
\title{Chiral-odd generalized parton distributions of sea quarks at $\xi=0$ in the light-cone quark model}
	
\author{Xiaoyan Luan}
\author{Zhun Lu}\email[]{zhunlu@seu.edu.cn}
\affiliation{School of Physics, Southeast University, Nanjing 211189, China}

\begin{abstract}
We study the chiral-odd generalized parton distributions (GPDs) of the $\bar{u}$ and $\bar{d}$ quarks inside the proton at zero skewness using the overlap representation within the light cone formalism. 
Using the light cone wave functions (LCWFs) of the proton obtained from the baryon-meson fluctuation model in terms of the $|q\bar{q}B\rangle$ Fock states, we provide the expressions of the GPDs $\widetilde{H}^{\bar{q}/P}_T(x,0,t)$, $H_T^{\bar{q}/P}(x,0,t)$ and $E_T^{\bar{q}/P}(x,0,t)$ for  $\bar{q}=\bar{u}$ and $\bar{d}$. Numerical results for these GPDs in momentum space as well as in impact parameter space are presented. We further investigate certain combinations of the chiral-odd GPDs in impact parameter space to the spin-orbit correlation effect of the sea quarks. 
\end{abstract}
	
\maketitle
	
\section{Introduction}\label{Sec1}
	     
Understanding the internal structure of hadrons in terms of constituent quarks, gluons and sea quarks is one of the main goals of QCD and hadronic physics. 
The generalized parton distributions (GPDs)~\cite{Muller:1994ses,Ji:1996nm,Radyushkin:1997ki,Diehl:2015uka}, usually viewed as the extension of the standard parton distribution functions (PDFs), have been recognized as important quantities describing the three-dimensional structure of the nucleon in addition to the transverse momentum dependent parton distributions (TMDs). 
The GPDs are off-forward matrix elements of nonlocal operators and can be accessed experimentally through the deeply virtual Compton scattering (DVCS)~\cite{Ji:1996nm,Radyushkin:1996nd,Belitsky:2001ns} or the	deeply virtual meson production (DVMP)~\cite{Goloskokov:2006hr,Goloskokov:2007nt,Goloskokov:2009ia,Goloskokov:2011rd}. 
At leading twist, there are eight GPDs: four chiral-even (quark helicity-non-flip) GPDs $H$,$E$,$\widetilde{H}$,$\widetilde{E}$ and  four chiral-odd (quark helicity-flip) GPDs $H_T$,$E_T$,$\widetilde{H}_T$,$\widetilde{E}_T$. The GPDs depend on three independent kinematic variables, the longitudinal momentum faction $x$ of the parton, the square of the total momentum transferred $t$ and the longitudinal momentum transferred skewness $\xi$. 
In the forward limit, the chiral-even GPDs and chiral-odd GPDs can reduce to the usual unpolarized and helicity-polarized quark PDFs and the quark transversity PDFs, respectively. 
On the one hand, the chiral-even GPDs encode richer knowledge on the orbital angular momentum (OAM) of quarks inside the nucleon~\cite{Radyushkin:1997ki,Ji:1996nm,Sehgal:1974rz,Kroll:2020jat}, electromagnetic and gravitational form factors~\cite{Brodsky:2000ii,Kumar:2014osa} as well as the charge and magnetization densities~\cite{Miller:2010nz,Kumar:2014coa}. 
On the other hand, the chiral-odd GPDs also provide information on the correlation between the spin and OMA carried by quarks inside the nucleon~\cite{Diehl:2005jf,Burkardt:2005hp}. 
Thus they contain a wealth of information about the partonic structure of the hadron. Through a Fourier transform with respect to the transverse momentum transfer $\boldsymbol{\Delta}_T$, one can obtain the distributions in the impact parameter space which provide tomographic description of the nucleon structure. Particularly, the impact parameter dependent GPDs have probabilistic interpretation and satisfy the positivity condition~\cite{Pisarski:2000eq,Burkardt:2002hr}.
	 	
In recent years, a lot of experimental and theoretical studies related to GPDs have been carried out. The experimental data based on the hard exclusive scattering were collected by the H1 collaboration~\cite{H1:1999pji,H1:2001nez,H1:2005gdw}, the ZEUS collaboration~\cite{ZEUS:1998xpo,ZEUS:2003pwh}, as well as the fixed target experiments at HERMES~\cite{HERMES:2001bob,HERMES:2011bou,HERMES:2012gbh}, COMPASS~\cite{dHose:2004usi} and JLab~\cite{CLAS:2001wjj}. 
The information of chiral-even GPDs can be obtained in the exclusive processes like DVCS~\cite{Ji:1996nm,Radyushkin:1996nd,Belitsky:2001ns} and hard exclusive meson production (HEMP)~\cite{Polyakov:1998ze,Collins:1996fb} on the basis of factorization theorems. 
Unlike the chiral-even GPDs, it is very difficult to measure chiral-odd GPDs because they have to combine with another chiral-odd object in the amplitude due to their helicity-flip character, otherwise they will decouple in most hard amplitudes. 
At present it is proposed that they can be accessed through deeply virtual pseudoscalar meson production processes which are particularly sensitive to chiral-odd GPDs~\cite{Ahmad:2008hp,Goloskokov:2011rd,Goldstein:2010gu}, either through the photon production of vector meson~\cite{Boussarie:2016qop}, or from the diffractive double meson production~\cite{Ivanov:2002jj, Enberg:2006he,Cosyn:2020kfe}. 
In a recent COMPASS measurement~\cite{COMPASS:2013fsk} on exclusive $\rho^{0}$ muonproduction by scattering muons off transversely polarized proton, a nonzero single-spin asymmetry $A_{UT}^{\sin\phi_S}$ was found. The data can be well described by a GPD-based model~\cite{Goloskokov:2013mba} using handbag approach, which is  interpreted as the first evidence for the existence of chiral-odd GPDs, especially the transversity GPD $H_T$. 
Furthermore, theoretical studies of hard exclusive pseudoscalar meson electroproduction~~\cite{Goloskokov:2011rd,Goldstein:2013gra,Diehl:2007hd,Duplancic:2016bge,Siddikov:2019ahb}, like $\pi^0$ and $\eta$ electroproduction~\cite{CLAS:2014jpc,CLAS:2017jjr,Kim:2015pkf,Goloskokov:2011rd,CLAS:2019uzc,Goldstein:2013gra,Ahmad:2008hp,Goldstein:2010gu}, indicate that the transversely polarized virtual photons have strong contributions that the transversity GPDs are essential in the description of the cross section. Therefore the contributions from chiral-odd GPDs should be considered in addition to the chiral-even GPDs. 
Notably, the simulations for the leading-twist contributions in $\gamma\rho$ photoproduction process from chiral-odd GPDs~\cite{Boussarie:2016qop,Beiyad:2010qg,Duplancic:2023kwe} are presently being performed~\cite{Pire:2019hos} in the future EIC within the kinematic range.
	
In this context, the chiral-odd GPDs have been investigated in various models. The first model calculation is the bag model, and only $H_T$ has been found to be non-zero~\cite{Scopetta:2005fg}. 
In Ref.~\cite{Pasquini:2005dk,Pincetti:2006hc}, the chiral-odd GPDs have been studied in a constituent quark model for nonzero skewness using the overlap representation in terms of LCWFs. In Ref.~\cite{Chakrabarti:2015ama}, they investigated the chiral-odd GPDs for both zero and nonzero skewness in the light front quark-diquark model predicted by the soft-wall AdS/QCD. The general properties of the chiral-odd GPDs have been investigated in transverse and longitudinal impact parameter spaces in Ref.~\cite{Chakrabarti:2008mw}. The impact parameter representation of the GPDs also have been studied in a QED model of a dressed electron~\cite{Dahiya:2007mt} and in a quark-diquark model~\cite{Kumar:2015yta} at zero skewness. In Refs.~\cite{Goldstein:2013gra,Goldstein:2014aja} ,the chiral-odd GPDs were studied using a physically motivated parameterization based on the reggeized diquark model.  The information about the Mellin moments of chiral odd GPDs has also been gained through the calculation starting from first principles as in lattice QCD~\cite{Gockeler:2005aw,Gockeler:2005cj,QCDSF-UKQCD:2008gss,Constantinou:2014fka,Lin:2017snn,Constantinou:2020hdm}. 
However most of those model calculation are about valence quarks, the knowledge of the sea quark Chiral-odd GPDs in a proton  is still limited.
	
In this paper, we apply the light-cone quark model to calculate the chiral-odd GPDs of the $\bar{u}$ and $\bar{d}$ quarks at zero skewedness using the overlap representation. We present the overlap representation in terms of LCWFs for the chiral-odd GPDs $H_T$, $E_T$, $\widetilde{H}_T$ and $\widetilde{E}_T$ in the general case. Then we calculate the chiral-odd GPDs of the sea quarks at $\xi=0$ where $\widetilde{E}_T$ does not contribute, as it is an odd function of $\xi$. To generate the sea quark degree freedom, we adopt the assumption proposed in Ref.~\cite{Brodsky:1996hc} that the proton can fluctuate to a composite state containing a meson $M$ and a baryon $B$, and $q\bar{q}$ are components of pion meson. And the LCWFs of the proton can be derived in terms of the $|q\bar{q}B\rangle$ Fock states which have been calculated in Ref. \cite{Luan:2022fjc}. In this framework, the chiral-odd GPDs of $\bar{u}$ and $\bar{d}$ can be obtained using those LCWFs. By taking Fourier transform (FT) with respect to $\boldsymbol{\Delta}_T$, the chiral-odd GPDs $H^{\bar{q}/P}$, $E^{\bar{q}/P}$, $\widetilde{H}^{\bar{q}/P}$ in impact parameter Space are also given. Using the chiral-odd GPDs in impact parameter space, we also show the numerical results of the certain combinations of chiral-odd GPDs $\mathcal{H}_{T}-\frac{\Delta_{b}}{4 m^{2}} \tilde{\mathcal{H}}_{T}$, $\mathcal{E}_{T}+2 \tilde{\mathcal{H}}_{T}$ and $\epsilon_{i j} b_{j} \frac{\partial}{\partial B}(\mathcal{E}_{T}+2 \tilde{\mathcal{H}}_{T})$.
	
The paper is organized as follows. In Sec. II, we derive the overlap representation in terms of LCWFs for the chiral-odd GPDs. In Sec. III, we apply LCWFs to calculate the chiral-old GPDs of the sea quarks.  In Sec. VI, we present the numerical results of the chiral-odd GPDs of the sea quarks in momentum as well as impact parameter space. We summarize the paper in Sec. V.
	  
\section{ Chiral-odd GPDs in overlap representation}\label{Sec2}
	 
The GPDs can be defined as the off-forward matrix elements of the quark-quark proton correlator function on the light cone:
\begin{align}\label{F0}
	  	F_{\Lambda^{\prime}, \Lambda}^{\Gamma}(x, \xi, t)=\left.\frac{1}{2} \int \frac{d z^{-}}{2 \pi} e^{i x \bar{P}^{+} z^{-}}\left\langle p^{\prime}, \Lambda^{\prime}\left|\bar{\psi}\left(-\frac{z}{2}\right) \Gamma \psi\left(\frac{z}{2}\right)\right| p, \Lambda\right\rangle\right|_{z^{+}=0, \mathbf{z}_{T}=0},
	  \end{align}
where $\Gamma$ is the Dirac matrix which can be chosen as $\gamma^{+}, \gamma^{+} \gamma_{5}, i \sigma^{i+} \gamma_{5} (i=1,2)$, and  $\Lambda, \Lambda^{\prime}$ denote the target helicities in the initial and final states. 
For the chiral-odd case, we take $\Gamma=i \sigma^{i+} \gamma_{5}$, thus $F_{\Lambda^{\prime}, \Lambda}^{i \sigma^{i+} \gamma_{5}}$ can be parameterized as~\cite{Rajan:2017cpx}
\begin{align}\label{F} 
	 \notag     F^{[i\sigma^{i+}\gamma^{5}]}_{{\Lambda^{\prime}}{\Lambda}}&=\frac{i \epsilon^{i j}}{2 P^{+}} \bar{U}\left(p^{\prime}, \Lambda^{\prime}\right)\left[i \sigma^{+j} H_{T}+\frac{\gamma^{+} \Delta^{j}-\Delta^{+} \gamma^{j}}{2 M} E_{T}+\frac{P^{+} \Delta^{j}}{M^{2}} \widetilde{H}_{T}-\frac{P^{+} \gamma^{j}}{M} \widetilde{E}_{T}\right] U(p, \Lambda)
	  	\notag \\&=\left[\frac{i \epsilon^{i j} \Delta^{j}}{2 M}\left(E_{T}+2 \tilde{H}_{T}\right)+\frac{\Lambda \Delta^{i}}{2 M}\left(\tilde{E}_{T}-\xi E_{T}\right)\right] \delta_{\Lambda^{\prime} \Lambda}+\left[\left(\delta_{i 1}+i \Lambda \delta_{i 2}\right) H_{T}-\frac{i \epsilon^{i j} \Delta^{j}\left(\Lambda \Delta^{1}+i \Delta^{2}\right)}{2 M^{2}} \tilde{H}_{T}\right] \delta_{-\Lambda^{\prime} \Lambda}.
\end{align}
Here, $\epsilon^{i j}$ is the antisymmetric tensor with $\epsilon^{1 2}=-\epsilon^{2 1}=1$,  $P=\left(p+p^{\prime}\right)/2$ is the average proton momentum, $\Delta=p^{\prime}-p$ is the momentum transfer to the proton with $t=\Delta^{2}=-{\Delta}_T^2$, and $\xi=-\Delta^{+}/2 P^{+}$ is the skewness parameter. 
      
We use $\uparrow$ ($\downarrow$) to denote the positive (negative) helicity of the proton. 
For $i=1$, we have
\begin{align}	F^1_{\uparrow\uparrow}&=\frac{i\Delta_2}{2M}(2\widetilde{H}_T+E_T)
+\frac{\Delta_1}{2M}(\widetilde{E}_T-\xi E_T)      	&F^1_{\downarrow\downarrow}&=\frac{i\Delta_2}{2M}(2\widetilde{H}_T+E_T)
-\frac{\Delta_1}{2M}(\widetilde{E}_T-\xi E_T),\label{eq:F1a}\\
 F^1_{\uparrow\downarrow}&=H_T+\frac{\widetilde{H}_T}{2M^2}(-i\Delta_2)(-\Delta_1+i\Delta_2)
 &F^1_{\downarrow\uparrow}&=H_T+\frac{\widetilde{H}_T}{2M^2}(-i\Delta_2)(\Delta_1+i\Delta_2).
 \label{eq:F1b}
\end{align}
While for $i=2$, we have
\begin{align}
F^2_{\uparrow\uparrow}&=\frac{-i\Delta_1}{2M}(2\widetilde{H}_T+E_T)
+\frac{\Delta_2}{2M}(\widetilde{E}_T-\xi E_T) 
&F^2_{\downarrow\downarrow}&=\frac{-i\Delta_1}{2M}(2\widetilde{H}_T+E_T)
-\frac{\Delta_2}{2M}(\widetilde{E}_T-\xi E_T),\label{eq:F2a}\\
F^2_{\uparrow\downarrow}&=-iH_T+\frac{\widetilde{H}_T}{2M^2}(i\Delta_1)(-\Delta_1+i\Delta_2)
&F^2_{\downarrow\uparrow}&=iH_T+\frac{\widetilde{H}_T}{2M^2}(i\Delta_1)(\Delta_1+i\Delta_2).
\label{eq:F2b}
\end{align}
Using Eqs.~(\ref{eq:F1a}, \ref{eq:F1b}, \ref{eq:F2a}, \ref{eq:F2b}), the Chiral-odd GPDs can be obtained from the following combinations: 
\begin{align} 	\frac{i\Delta_1\Delta_2}{M^2}\widetilde{H}_T=\frac{F^1_{\uparrow\downarrow}
-F^1_{\downarrow\uparrow}}{2}-\frac{i(F^2_{\uparrow\downarrow}+F^2_{\downarrow\uparrow})}{2},   \\
2H_T+\frac{\Delta_T^2}{2M^2}\widetilde{H}_T=\frac{F^1_{\uparrow\downarrow}
+F^1_{\downarrow\uparrow}}{2}+\frac{i(F^2_{\uparrow\downarrow}
-F^2_{\downarrow\uparrow})}{2},\\
\frac{\Delta_1+i\Delta_2}{2M}(\widetilde{E}_T-\xi E_T)=\frac{F^1_{\uparrow\uparrow}-F^1_{\downarrow\downarrow}}{2}
+\frac{i(F^2_{\uparrow\uparrow}-F^2_{\downarrow\downarrow})}{2},\\
\frac{\Delta_1+i\Delta_2}{2M}(2\widetilde{H}_T+E_T)=\frac{F^1_{\uparrow\uparrow}+F^1_{\downarrow\downarrow}}{2}+\frac{i(F^2_{\uparrow\uparrow}+F^2_{\downarrow\downarrow})}{2}.      	
\end{align}
  
According to Ref.~\cite{Diehl:2001pm}, the GPDs can be related to the following matrix elements
     \begin{align}
     	A_{\Lambda^{\prime} \mu^{\prime}, \Lambda \mu}= \left.\int \frac{d z^{-}}{2 \pi} e^{i x P^{+} z^{-}}\left\langle p^{\prime}, \Lambda^{\prime}\left|\mathcal{O}_{\mu^{\prime}, \mu}(z)\right| p, \Lambda\right\rangle\right|_{z^{+} = 0, \mathbf{z}_{T}= 0},
     \end{align}
where $\mu^{\prime}$ and $\mu$ denote the helicities of the active parton. The operators $\mathcal{O}_{\mu^{\prime}, \mu}$ occurring in the definitions of the quark distributions have been given in Ref.~\cite{Diehl:2001pm}. 
Similarly, the case of antiquark can be written as
     \begin{align}
     	\mathcal{O}_{-,+} & = \frac{i}{4} \psi \sigma^{+1}\left(1-\gamma_{5}\right) \bar{\psi} = -\frac{i}{4} \bar{\psi} \sigma^{+1}\left(1-\gamma_{5}\right) \psi ,\\
     	\mathcal{O}_{+,-} & = -\frac{i}{4} \psi \sigma^{+1}\left(1+\gamma_{5}\right) \bar{\psi}= \frac{i}{4} \bar{\psi} \sigma^{+1}\left(1+\gamma_{5}\right) \psi.
     \end{align}
Here, $+(-)$ denotes the positive (negative) helicity of the antiquark, which is different from the case for antiquark in Ref.~\cite{Diehl:2005jf} where $+(-)$ donetes the quark helicity. Compared to the case of quarks, there is a global negative sign in the case of antiquarks because the order of the operators $\bar{\psi}$ and $\psi$ has to be reversed to obtain a density operator for antiquarks. 
The correlation functions in Eq.(\ref{F}) thus can be written in terms of the antiquark-proton helicity amplitudes as
     \begin{align}
        F^1_{{\Lambda^{\prime}}{\Lambda}} & =-( A_{\Lambda^{\prime}+, \Lambda-}+A_{\Lambda^{\prime}-, \Lambda+}) ,\\
    	F^2_{{\Lambda^{\prime}}{\Lambda}}& =i (A_{\Lambda^{\prime}-, \Lambda+}-A_{\Lambda^{\prime}+, \Lambda-}) .
     \end{align}
Here, the relation $\Gamma=\sigma^{i+}=-\epsilon^{ij} i \sigma^{j+} \gamma_{5}$ is used.
   	
Within the light-cone approach, the Fock-state expansion for a proton can be written as 
   \begin{align}
   	\notag \left|p, \Lambda\right\rangle  = \sum_{n} \prod_{i= 1}^{n} \frac{\mathrm{d} x_{i} \mathrm{~d}^{2} k_\perp ^i}{\sqrt{x_{i}} 16 \pi^{3}} 16\pi^3\delta(1-\sum_{j}x_j)\delta^2(\sum_{j=1}^{n}k^j_\bot)
   	 \psi_{n}\left(x_{i}, k_\perp ^i, \lambda_{i}\right)\left|n ; x_{i} p^{+}, x_{i} p_{\perp}+k_\perp ^i, \lambda_{i}\right\rangle .
   \end{align}
Similar to the case of the chiral-even GPDs~\cite{Brodsky:2000xy}, there are also contributions from the $n\to n$ diagonal overlap in the kinematical region $\xi<x<1$ and $\xi-1<x<0$. 
Therefore, Eq.~(\ref{F}) can be expressed through the overlap representation in terms of the LCWFs as follows~\cite{Chakrabarti:2008mw}:
      \begin{align}
      	\notag	  F^1_{{\Lambda^{\prime}}{\Lambda}}=&-(1-\xi)^{(1-\frac{n}{2})}
      \sum_{{\lambda_i},n}\int\prod_{i=1}^{n}\frac{dx_id^2k^i_\bot}
      {16\pi^3}16\pi^3\delta(1-\sum_{j}x_j)\delta^2(\sum_{j=1}^{n}k^j_\bot)\delta(x-x_1)
      	\\&\times \psi^{\Lambda^{\prime}*}_n(x_i^\prime,k_\bot^{\prime i},,\lambda_i^\prime)\psi^{\Lambda}_n(x_i,k_\bot^{ i},\lambda_i)\delta_{\lambda_1^\prime,-\lambda_1}[\delta_{\lambda_i^\prime,\lambda_i}
      (i=2...n)],\\
      	\notag	  F^2_{{\Lambda^{\prime}}{\Lambda}}=&i(1-\xi)^{(1-\frac{n}{2})}
      \sum_{{\lambda_i},n}\int\prod_{i=1}^{n}\frac{dx_id^2k^i_\bot}{16\pi^3}
      16\pi^3\delta(1-\sum_{j}x_j)\delta^2(\sum_{j=1}^{n}k^j_\bot)\delta(x-x_1)
      	\\&\times sign(\lambda_1)\psi^{\Lambda^{\prime}*}_n(x_i^\prime,k_\bot^{\prime i},,\lambda_i^\prime)\psi^{\Lambda}_n(x_i,k_\bot^{ i},\lambda_i)\delta_{\lambda_1^\prime,-\lambda_1}
      [\delta_{\lambda_i^\prime,\lambda_i}(i=2...n)],
      \end{align}
where $\lambda_1(\lambda_1^\prime)$  represents the  helicity of the initial(final) struck antiquark, and $\lambda_i(\lambda_i^\prime)$ denotes the helicity of the initial(final) spectators.
      
We thus obtain the formulae for the  chiral-odd GPDs within the overlap representation in terms of the proton LCWFs
      \begin{align}\label{widetilde{H}_T}
      	\notag
      	\frac{i\Delta_1\Delta_2}{M^2}\widetilde{H}_T(x, \xi, t)=&-(1-\xi)^{(1-\frac{n}{2})}\sum_{{\lambda_i},n}\int\prod_{i=1}^{n}
      \frac{dx_id^2k^i_\bot}{16\pi^3}16\pi^3\delta(1-\sum_{j}x_j)
      \delta^2(\sum_{j=1}^{n}k^j_\bot)\delta(x-x_1)
      	\\&\times\left[ \psi^{\uparrow*}_{+n}(x_i^\prime,k_\bot^{\prime i},\lambda_i^\prime)\psi^{\downarrow}_{-n}(x_i,k_\bot^{ i},\lambda_i)-\psi^{\downarrow*}_{-n}(x_i^\prime,k_\bot^{\prime i},\lambda_i^\prime)\psi^{\uparrow}_{+n}(x_i,k_\bot^{ i},\lambda_i)\right](i=2...n),
      \displaybreak[0]\\
      \label{bar{H}_T}
      	\notag
      	2H_T+\frac{\Delta_T^2}{2M^2}\widetilde{H}_T(x, \xi, t)=&-(1-\xi)^{(1-\frac{n}{2})}\sum_{{\lambda_i},n}\int\prod_{i=1}^{n}
      \frac{dx_id^2k^i_\bot}{16\pi^3}16\pi^3\delta(1-\sum_{j}x_j)
      \delta^2(\sum_{j=1}^{n}k^j_\bot)\delta(x-x_1)
      	\\&\times\left[ \psi^{\uparrow*}_{-n}(x_i^\prime,k_\bot^{\prime i},\lambda_i^\prime)\psi^{\downarrow}_{+n}(x_i,k_\bot^{ i},\lambda_i)+\psi^{\downarrow*}_{+n}(x_i^\prime,k_\bot^{\prime i},\lambda_i^\prime)\psi^{\uparrow}_{-n}(x_i,k_\bot^{ i},\lambda_i)\right](i=2...n),\displaybreak[0]\\
\label{widetilde{E}_T}
      	\notag
      	\frac{\Delta_1+i\Delta_2}{2M}(\widetilde{E}_T-\xi E_T)(x, \xi, t)=&-(1-\xi)^{(1-\frac{n}{2})}\sum_{{\lambda_i},n}\int\prod_{i=1}^{n}
      \frac{dx_id^2k^i_\bot}{16\pi^3}16\pi^3\delta(1-\sum_{j}x_j)
      \delta^2(\sum_{j=1}^{n}k^j_\bot)\delta(x-x_1)
      	\\&\times\left[ \psi^{\uparrow*}_{-n}(x_i^\prime,k_\bot^{\prime i},\lambda_i^\prime)\psi^{\uparrow}_{+n}(x_i,k_\bot^{ i},\lambda_i)-\psi^{\downarrow*}_{-n}(x_i^\prime,k_\bot^{\prime i},\lambda_i^\prime)\psi^{\downarrow}_{+n}(x_i,k_\bot^{ i},\lambda_i)\right](i=2...n),\displaybreak[0]\\
\label{bar{E}_T}
      	\notag
      	\frac{\Delta_1+i\Delta_2}{2M}(2\widetilde{H}_T+E_T)(x, \xi, t)=&-(1-\xi)^{(1-\frac{n}{2})}\sum_{{\lambda_i},n}\int\prod_{i=1}^{n}
      \frac{dx_id^2k^i_\bot}{16\pi^3}16\pi^3\delta(1-\sum_{j}x_j)
      \delta^2(\sum_{j=1}^{n}k^j_\bot)\delta(x-x_1)\displaybreak[0]
      	\\&\times\left[ \psi^{\uparrow*}_{-n}(x_i^\prime,k_\bot^{\prime i},\lambda_i^\prime)\psi^{\uparrow}_{+n}(x_i,k_\bot^{ i},\lambda_i)+\psi^{\downarrow*}_{-n}(x_i^\prime,k_\bot^{\prime i},\lambda_i^\prime)\psi^{\downarrow}_{+n}(x_i,k_\bot^{ i},\lambda_i)\right](i=2...n).\displaybreak[0]
      \end{align}
      
      \section{ chiral-odd GPDs of the sea quarks }\label{Sec2}
In this section, we present the calculation on the chiral-odd GPDs of the $\bar{u}$ and $\bar{d}$ quarks in the proton at zero skewness in the light-cone quark model. 
On the one hand, the light-cone formalism has been widely applied in the calculation of parton distribution functions of nucleon and meson~\cite{Lepage:1980fj}. 
Within the light-cone approach, the wave functions for a hadronic composite state can be expressed as LCWFs in Fock-state basis. On the other hand, the overlap representation has also been used to study various form factors of the hadrons~\cite{Brodsky:2000ii} and the pion~\cite{Xiao:2003wf}, anomalous magnetic moment of the nucleon~\cite{Brodsky:2000ii} as well as GPDs~\cite{Brodsky:2000xy}. Here we extend light-cone formalism to calculate the chiral-odd GPDs of the sea quarks. 
     
In the light-cone approach, the wave functions of the hadron, which describe a hadronic composite state at a particular light-cone time, are expressed in terms of a series of LCWFs in the Fock-state basis.
     
In order to  generate the sea quark degree of freedom, we apply the baryon-meson fluctuation model~\cite{Brodsky:1996hc}, in which the proton can fluctuate to a composite system formed by a meson $M$ and a baryon $B$, where the meson is composed in terms of $q\bar{q}$:
     \begin{align}\label{fock state}
     	|p\rangle\to| M B\rangle\to|q\bar{q}B\rangle.
     \end{align}
     The full LCWFs have been derived in Ref.~\cite{Luan:2022fjc} and have the forms
     \begin{align}\label{LCWFs}
     	\psi^{\lambda_N}_{{\lambda_B}{\lambda_q}{\lambda_{\bar{q}}}}
     	(x,y,\boldsymbol{k}_T,\boldsymbol{r}_T)
=&\psi^{\lambda_N}_{\lambda_B}(y,\boldsymbol{r}_T)\psi_{{\lambda_q}{\lambda_{\bar{q}}}}
     	(x,y,\boldsymbol{k}_T,\boldsymbol{r}_T),
     \end{align}
where $\psi^{\lambda_N}_{\lambda_B}(y,r_T)$ can be viewed as the wave function of the nucleon in terms of the $\pi B$ components, and $\psi_{{\lambda_q}{\lambda_{\bar{q}}}}
(x,y,\boldsymbol{k}_T,\boldsymbol{r}_T)$ is the  wave function of the pion in terms of the $q \bar{q}$ components. 
The indices $\lambda_N$, $\lambda_B$, $\lambda_q$ and $\lambda_{\bar{q}}$ denote the helicity of the proton, baryon, quark and antiquark, respectively. 
Finally, $x$ and $y$ represent the light-cone momentum fractions, while $\boldsymbol{k}_T$ and $\boldsymbol{r}_T$ denote the transverse momenta of the antiquark and the meson. 
     
For $\psi^{\lambda_N}_{\lambda_B}(y,r_T)$ in Eq.~(\ref{LCWFs}), they have the expressions:
     \begin{align}\label{former}
     	\notag\psi^+_+(y,\boldsymbol{r}_T)&=\frac{M_B-(1-y)M}{\sqrt{1-y}}\phi_1, \\
     	\notag\psi^+_-(y,\boldsymbol{r}_T)&=\frac{r_1+ir_2}{\sqrt{1-y}}\phi_1, \\
     	\notag\psi^-_+(y,\boldsymbol{r}_T)&=\frac{r_1-ir_2}{\sqrt{1-y}}\phi_1 , \\
     	\psi^-_-(y,\boldsymbol{r}_T)&=\frac{(1-y)M-M_B}{\sqrt{1-y}}\phi_1.
     \end{align}	
Here, $M$ and $M_B$ are the masses of proton and baryon, respectively. 
$\phi_1$ is the wave function of the baryon-meson system in the momentum space with the form
     \begin{align}	\phi_1(y,\boldsymbol{r}_T)=-\frac{g(r^2)\sqrt{y(1-y)}}{\boldsymbol{r}_T^2+L_1^2(m_\pi^2)},
     \end{align}
     where $m_\pi$ is the mass of $\pi$ meson, $g(r^2)$ is the form factor for the coupling of the nucleon-pion meson-baryon vertex, and
     \begin{align}
     	L_1^2({m_\pi^2})=yM_B^2+(1-y){m_\pi^2}-y(1-y)M^2.
     \end{align}
     
The pion LCWFs in Eq.~(\ref{LCWFs}) have the following expressions:
     \begin{align}\label{later}	\notag\psi{_+}{_+}(x,y,\boldsymbol{k}_T,\boldsymbol{r}_T)&=
     \frac{my}{\sqrt{x(y-x)}}\phi_2,\\	\notag\psi{_+}{_-}(x,y,\boldsymbol{k}_T,\boldsymbol{r}_T)&=
     	\frac{y(k_1-ik_2)-x(r_1-ir_2)}{\sqrt{x(y-x)}}\phi_2, \\      	\notag\psi{_-}{_+}(x,y,\boldsymbol{k}_T,\boldsymbol{r}_T)&=
     	\frac{y(k_1+ik_2)-x(r_1+ir_2)}{\sqrt{x(y-x)}}\phi_2, \\   	\psi{_-}{_-}(x,y,\boldsymbol{k}_T,\boldsymbol{r}_T)&=\frac{-my}{\sqrt{x(y-x)}}\phi_2,
     \end{align} 
     Here, $m$ is the mass of quarks and sea quarks, and
     \begin{align}	     	 	\phi_2(x,y,\boldsymbol{k}_T,\boldsymbol{r}_T)=-\frac{g(k^2)\sqrt{\frac{x}{y}(1-\frac{x}{y})}}
     	{(\boldsymbol{k}_T-\frac{x}{y}\boldsymbol{r}_T)^2+L_2^2(m^2)},
     \end{align}
is the wavefunction of the pion meson in the momentum space, with  
     \begin{align}	L_2^2(m^2)=\frac{x}{y}m^2+\left(1-\frac{x}{y}\right)m^2-\frac{x}{y}
     	\left(1-\frac{x}{y}\right){m_\pi}^2.
     \end{align}
and $g(k^2)$ is the form factor for the coupling of the pion meson-quark-sea quark vertex.
In this work, we adopt $g(r^2)$ and $g(k^2)$ as the dipolar form factor
     \begin{align}
     	g(r^2)&=-g_1(1-y)\frac{\boldsymbol{r}_T^2+L_1^2(m_\pi^2)}
     	{[\boldsymbol{r}_T^2+L_1^2(\Lambda^2_\pi)]^2},\label{eq15}\\ 	g(k^2)&=-g_2(1-\frac{x}{y})\frac{(\boldsymbol{k}_T-\frac{x}{y}
     		\boldsymbol{r}_T)^2+L_2^2(m^2)}{[(\boldsymbol{k}_T-\frac{x}{y}
     		\boldsymbol{r}_T)^2+L_2^2(\Lambda^2_{\bar{q}})]^2}.\label{eq16}	
     \end{align}
     
Using the overlap representation in Eqs.~(\ref{widetilde{H}_T}-\ref{bar{E}_T}), the chiral-odd GPDs for sea quarks can be calculated from:
\begin{align}	\notag\frac{i\Delta_1\Delta_2}{M^2}\widetilde{H}_T=&-\sum_{{\lambda_B}{\lambda_q}}
\int\frac{d^2\boldsymbol{k}_T}{16\pi^3}\int\frac{d^2\boldsymbol{r}_T}{16\pi^3}	\\
&\left[\psi^{\uparrow*}_{{\lambda_B}{\lambda_q}+}(x,y,\boldsymbol{k}^{\prime}_T,
\boldsymbol{r}^{\prime}_T)\psi^{\downarrow}_{{\lambda_B}{\lambda_q}-}
(x,y,\boldsymbol{k}_T,\boldsymbol{r}_T)-\psi^{\downarrow*}_{{\lambda_B}{\lambda_q}-}
(x,y,\boldsymbol{k}^{\prime}_T,\boldsymbol{r}^{\prime}_T)\psi^{\uparrow}_{{\lambda_B}
{\lambda_q}+}(x,y,\boldsymbol{k}_T,\boldsymbol{r}_T)\right],\\
\notag 2H_T+\frac{\Delta_T^2}{2M^2}\widetilde{H}_T=&-\sum_{{\lambda_B}{\lambda_q}}
\int\frac{d^2\boldsymbol{k}_T}{16\pi^3}\int\frac{d^2\boldsymbol{r}_T}{16\pi^3}
\\&\left[\psi^{\downarrow*}_{{\lambda_B}{\lambda_q}+}(x,y,\boldsymbol{k}^{\prime}_T,
\boldsymbol{r}^{\prime}_T)\psi^{\uparrow}_{{\lambda_B}{\lambda_q}-}
(x,y,\boldsymbol{k}_T,\boldsymbol{r}_T)+\psi^{\uparrow*}_{{\lambda_B}{\lambda_q}-}
(x,y,\boldsymbol{k}^{\prime}_T,\boldsymbol{r}^{\prime}_T)
\psi^{\downarrow}_{{\lambda_B}{\lambda_q}+}(x,y,\boldsymbol{k}_T,\boldsymbol{r}_T)\right],
      \\ 	\notag\frac{\Delta_1+i\Delta_2}{2M}\widetilde{E}_T=&-\sum_{{\lambda_B}{\lambda_q}}
      \int\frac{d^2\boldsymbol{k}_T}{16\pi^3}\int\frac{d^2\boldsymbol{r}_T}{16\pi^3}		\\
      &\left[\psi^{\uparrow*}_{{\lambda_B}{\lambda_q}-}(x,y,\boldsymbol{k}^{\prime}_T,
      \boldsymbol{r}^{\prime}_T)\psi^{\uparrow}_{{\lambda_B}{\lambda_q}+}
      (x,y,\boldsymbol{k}_T,\boldsymbol{r}_T)-\psi^{\downarrow*}_{{\lambda_B}{\lambda_q}-}
      (x,y,\boldsymbol{k}^{\prime}_T,\boldsymbol{r}^{\prime}_T)
      \psi^{\downarrow}_{{\lambda_B}{\lambda_q}+}(x,y,\boldsymbol{k}_T,\boldsymbol{r}_T)
      \right],
    \\
      	\notag\frac{\Delta_1+i\Delta_2}{2M}(2\widetilde{H}_T+E_T)=
      &-\sum_{{\lambda_B}{\lambda_q}}\int\frac{d^2\boldsymbol{k}_T}
      {16\pi^3}\int\frac{d^2\boldsymbol{r}_T}{16\pi^3}  	\\
      &\left[\psi^{\uparrow*}_{{\lambda_B}{\lambda_q}-}(x,y,\boldsymbol{k}^{\prime}_T,
      \boldsymbol{r}^{\prime}_T)\psi^{\uparrow}_{{\lambda_B}{\lambda_q}+}(x,y,
      \boldsymbol{k}_T,\boldsymbol{r}_T)+\psi^{\downarrow*}_{{\lambda_B}{\lambda_q}-}
      (x,y,\boldsymbol{k}^{\prime}_T,\boldsymbol{r}^{\prime}_T)
      \psi^{\downarrow}_{{\lambda_B}{\lambda_q}+}(x,y,\boldsymbol{k}_T,\boldsymbol{r}_T)
      \right].
      \end{align}
where 
       \begin{align}	\nonumber\boldsymbol{k}^{{\prime}{\prime}}_T&=\boldsymbol{k}_T
       -\frac{1}{2}(1-x)\boldsymbol{\Delta}_T  \\
       \boldsymbol{k}^{\prime}_T&=\boldsymbol{k}_T+\frac{1}{2}(1-x)\boldsymbol{\Delta}_T,
       \end{align}	
are the transverse momenta for the final-state and initial-state struck antiquarks,       \begin{align}	\nonumber-\boldsymbol{r}^{{\prime}{\prime}}_T&=-\boldsymbol{r}_T
       +\frac{1}{2}(1-y)\boldsymbol{\Delta}_T   \\\nonumber-\boldsymbol{r}^{\prime}_T&=-\boldsymbol{r}_T-\frac{1}{2}(1-y)
       \boldsymbol{\Delta}_T
       	\\\nonumber(\boldsymbol{r}_T-\boldsymbol{k}_T)^{{\prime}{\prime}}&=	(\boldsymbol{r}_T-\boldsymbol{k}_T)+\frac{1}{2}(y-x)\boldsymbol{\Delta}_T 
       	\\(\boldsymbol{r}_T-\boldsymbol{k}_T)^{\prime}&=	(\boldsymbol{r}_T-\boldsymbol{k}_T)-\frac{1}{2}(y-x)\boldsymbol{\Delta}_T,
       \end{align}	
are the transverse momenta for the final and initial spectators $B$ and  $q$, respectively.

Substituting the light-cone wave functions of the proton in Eq.~(\ref{former})(\ref{later}), we obtain the expressions for the chiral-odd GPDs of the sea quarks as follows:
       \begin{align}
       \widetilde{E}_T^{\overline{q}/P}(x,0,t)&=0, \\
       	H_T^{\overline{q}/P}(x,0,t)&=0,\\
\widetilde{H}_T^{\overline{q}/P}(x,0,t)&=\frac{g_1^2g_2^2}{(2\pi)^6}
\int_{x}^{1}\frac{dy}{y}\int d^2\boldsymbol{k}_T\int{d^2\boldsymbol{r}_T}\frac{y(1-y)^3(1-\frac{x}{y})^3M^2m[M_B-(1-y)M]}{D_1(y,\boldsymbol{r}_T,\boldsymbol{\Delta}_T) D_2(\frac{x}{y},\boldsymbol{k}_T-\frac{x}{y}\boldsymbol{r}_T,\boldsymbol{\Delta}_T)},\\       	E_T^{\overline{q}/P}(x,0,t)&=\frac{g_1^2g_2^2}{(2\pi)^6}\int_{x}^{1}\frac{dy}{y}\int d^2\boldsymbol{k}_T\int{d^2\boldsymbol{r}_T}
       	\\&	\frac{y(1-y)^2(1-\frac{x}{y})^3\left\{Mm\left[[M_B-(1-y)M]^2+\boldsymbol{r}_T^2-\frac{1}{4}(1-y)\boldsymbol{\Delta}_T^2\right]-2M^2m(1-y)[M_B-(1-y)M]\right\}}{D_1(y,\boldsymbol{r}_T,\boldsymbol{\Delta}_T) D_2(\frac{x}{y},\boldsymbol{k}_T-\frac{x}{y}\boldsymbol{r}_T,\boldsymbol{\Delta}_T)}.
       \end{align}
       where
      \begin{align}      	 D_1(y,\boldsymbol{r}_T,\boldsymbol{\Delta}_T)&=\left[(\boldsymbol{r}_T-\frac{1}{2}(1-y)
      \boldsymbol{\Delta}_T)^2+L_1^2\right]^2 \left[(\boldsymbol{r}_T+\frac{1}{2}(1-y)\boldsymbol{\Delta}_T)^2+L_1^2\right]^2,\\
      D_2(\frac{x}{y},\boldsymbol{k}_T-\frac{x}{y}\boldsymbol{r}_T,\boldsymbol{\Delta}_T)
      &=\left[[(\boldsymbol{k}_T-\frac{x}{y}\boldsymbol{r}_T)-\frac{1}{2}(1-\frac{x}{y})
      \boldsymbol{\Delta}_T]^2+L_2^2\right]^2
         \left[[(\boldsymbol{k}_T-\frac{x}{y}\boldsymbol{r}_T)
         +\frac{1}{2}(1-\frac{x}{y})\boldsymbol{\Delta}_T]^2+L_2^2\right]^2.
      \end{align}
      Our results show that two chiral-odd GPDs of the antiquarks  , $\widetilde{E}_T^{\overline{q}/P}(x,0,t)$ and
       	$H_T^{\overline{q}/P}(x,0,t)$ are zero.
Here, $\widetilde{E}_T$ does not contribute at $\xi=0$ because it is an odd function of $\xi$, which is consistent with our model calculation result. 
As for $H_T$, it reduces to the transversity distribution $h_1$ in the forward limit. 
The sea quark transversity distributions are usually assumed to be zero in many analyses due to the fact that quark transversity distributions do not mix with gluons in the evolution. In a recent phenomenological extraction of transversity distribution functions by simultaneously fitting to semi-inclusive deep inelastic scattering and electronpositron annihilation data~\cite{Zeng:2023nnb}, it was found that the $\bar{u}$ quark favors a negative transversity distribution while that of the $\bar{d}$ quark is consistent with zero according to the current accuracy. 
The reason for the difference may be that in our model calculation, we include $\bar{u}$ and $\bar{d}$ flavors in one expression, and thus the function is less constrained.

\section{Numerical results for chiral-odd GPDs of sea quarks }\label{Sec3}
In this section, we present the numerical results for the chiral-odd GPDs of the sea quarks in momentum as well as impact parameter space. To present the numerical results of the sea quark chiral-odd GPDs, we need to specify the values of the parameters in our model. We choose the values from Ref.~\cite{Luan:2022fjc}:
      \begin{center}\label{tab1}
      	\setlength{\tabcolsep}{5mm}
      	\renewcommand\arraystretch{1.5}
      	\begin{tabular}{ c | c | c }
      		\hline
      		Parameters & $\bar{u}$ & $\bar{d}$ \\
      		\hline
      		\hline
      		$g_1$ & 9.33 & 5.79 \\
      		\hline
      		$g_2$ & 4.46 & 4.46 \\
      		\hline
      		$\Lambda_\pi(GeV)$ &  0.223 & 0.223 \\
      		\hline
      		$\Lambda_{\bar{q}}(GeV)$ &  0.510 &  0.510 \\
      		\hline
      	\end{tabular}
      	\captionof{table}{Values of the parameters taken from Ref.~\cite{Luan:2022fjc}.} \label{tab1}
      \end{center}
As shown in Ref.~\cite{Luan:2022fjc}, the values of $g_2$ and $\Lambda_\pi$ are fixed by adopting the GRV leading-order (LO) parametrization~\cite{Gluck:1991ey} to perform the fit for $f_1^{\bar{u}/\pi^-}$ (or $f_1^{\bar{d}/\pi^+}(x)$). The MSTW2008 LO parametrization~\cite{Martin:2009iq} are adopted for $f_1^{\bar{u}/P}$ and $f_1^{\bar{d}/P}$ to obtain the values of the parameters $g_1$ and $\Lambda_{\bar{q}}$.
      
      \begin{figure*}[htbp]
      	\centering
      	\subfigure{\begin{minipage}[b]{0.45\linewidth}
      			\centering
      			\includegraphics[width=\linewidth]{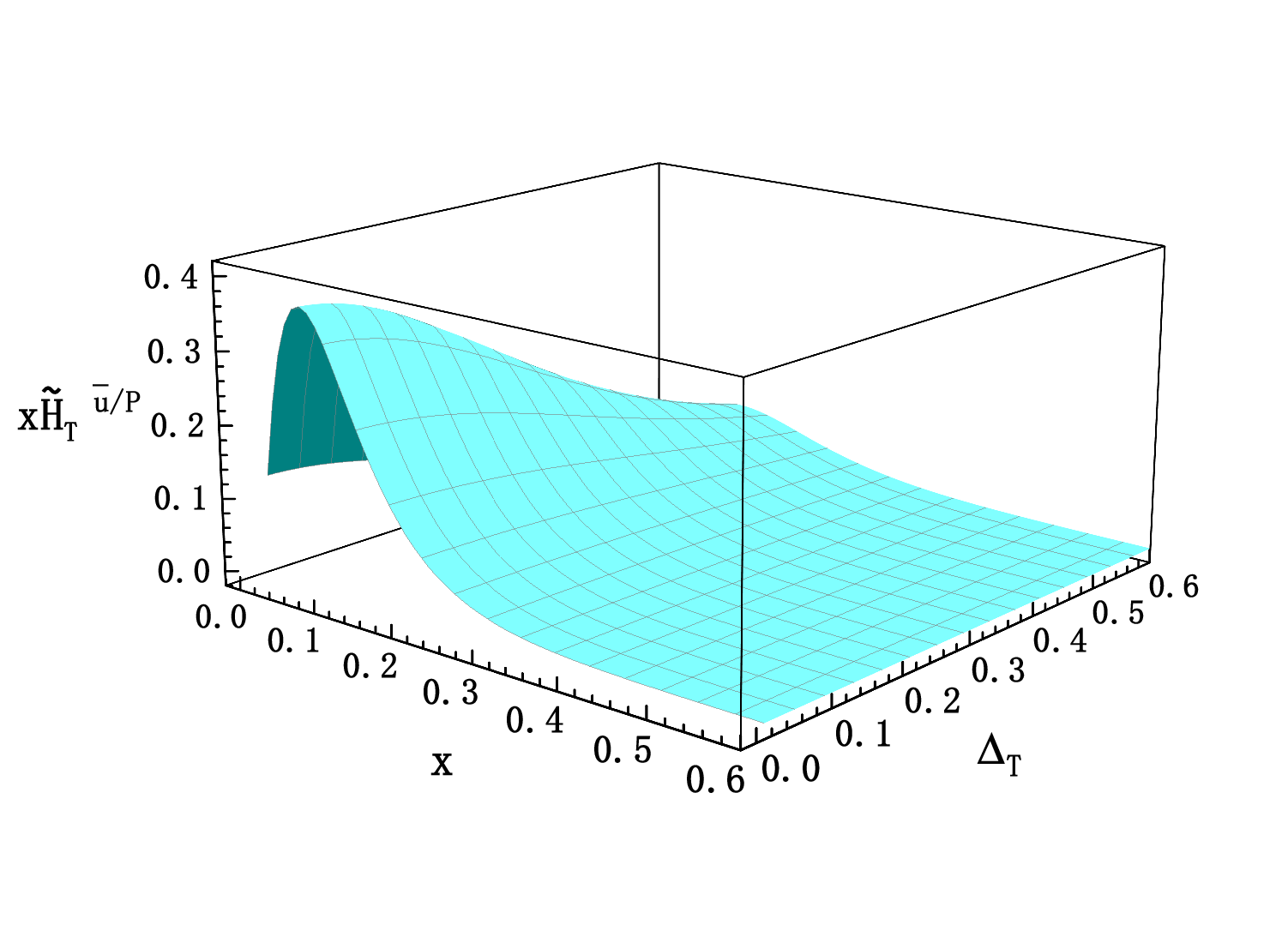}
      	\end{minipage}}
      	\subfigure{\begin{minipage}[b]{0.45\linewidth}
      			\centering
      			\includegraphics[width=\linewidth]{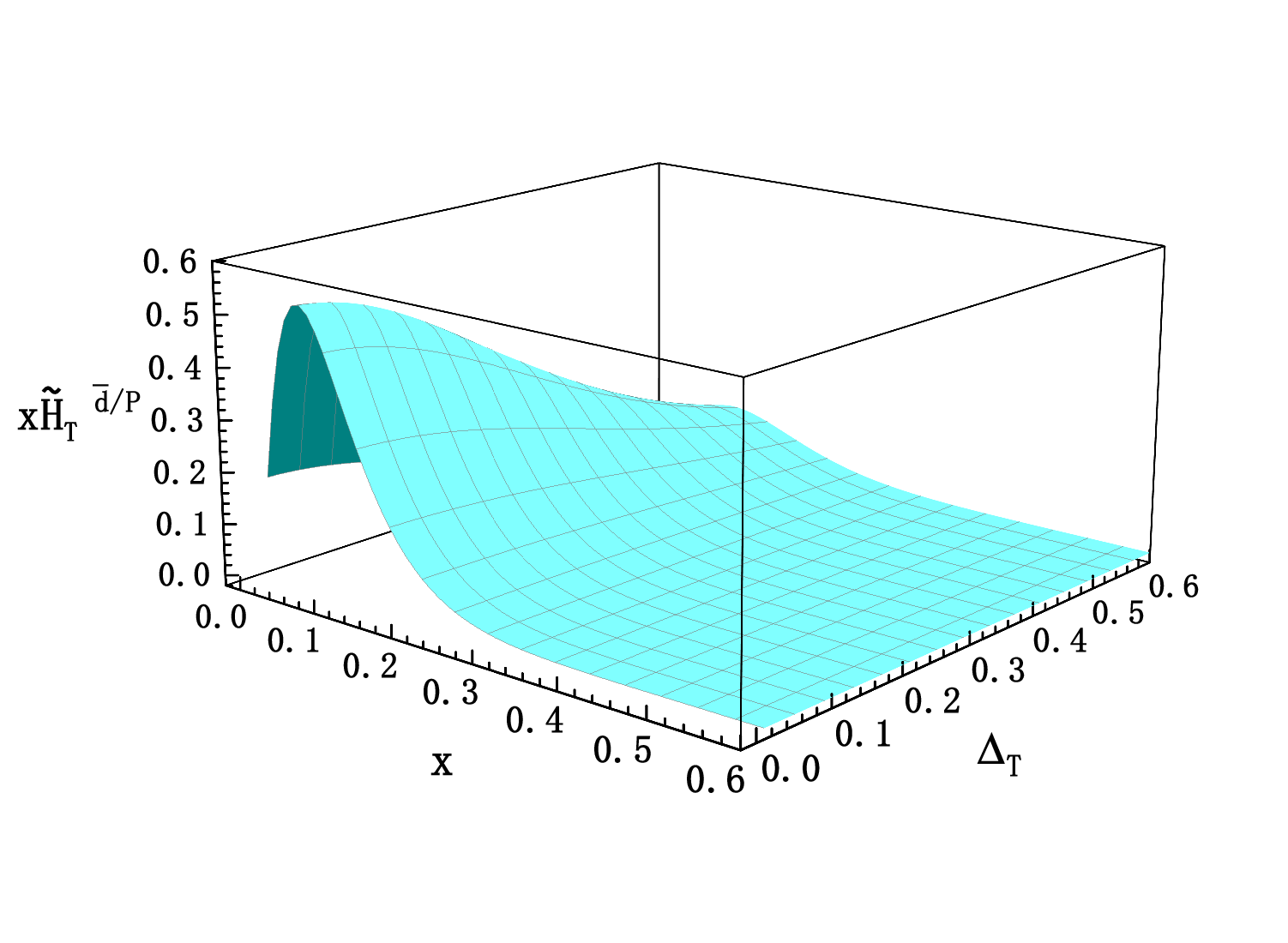}    	\end{minipage}}
      	\caption{The chiral-odd GPDs  $\widetilde{H}^{\bar{u}/P}_T(x,0,-\boldsymbol{\Delta}_T^2)$ and $\widetilde{H}^{\bar{d}/P}_T(x,0,-\boldsymbol{\Delta}_T^2)$ in the light-cone quark model as functions of $x$ and $\Delta_T$.} \label{Ht1}      
      \end{figure*}
      \begin{figure*}[htbp]
      	\centering
      	\subfigure{\begin{minipage}[b]{0.45\linewidth}
      			\centering
      			\includegraphics[width=\linewidth]{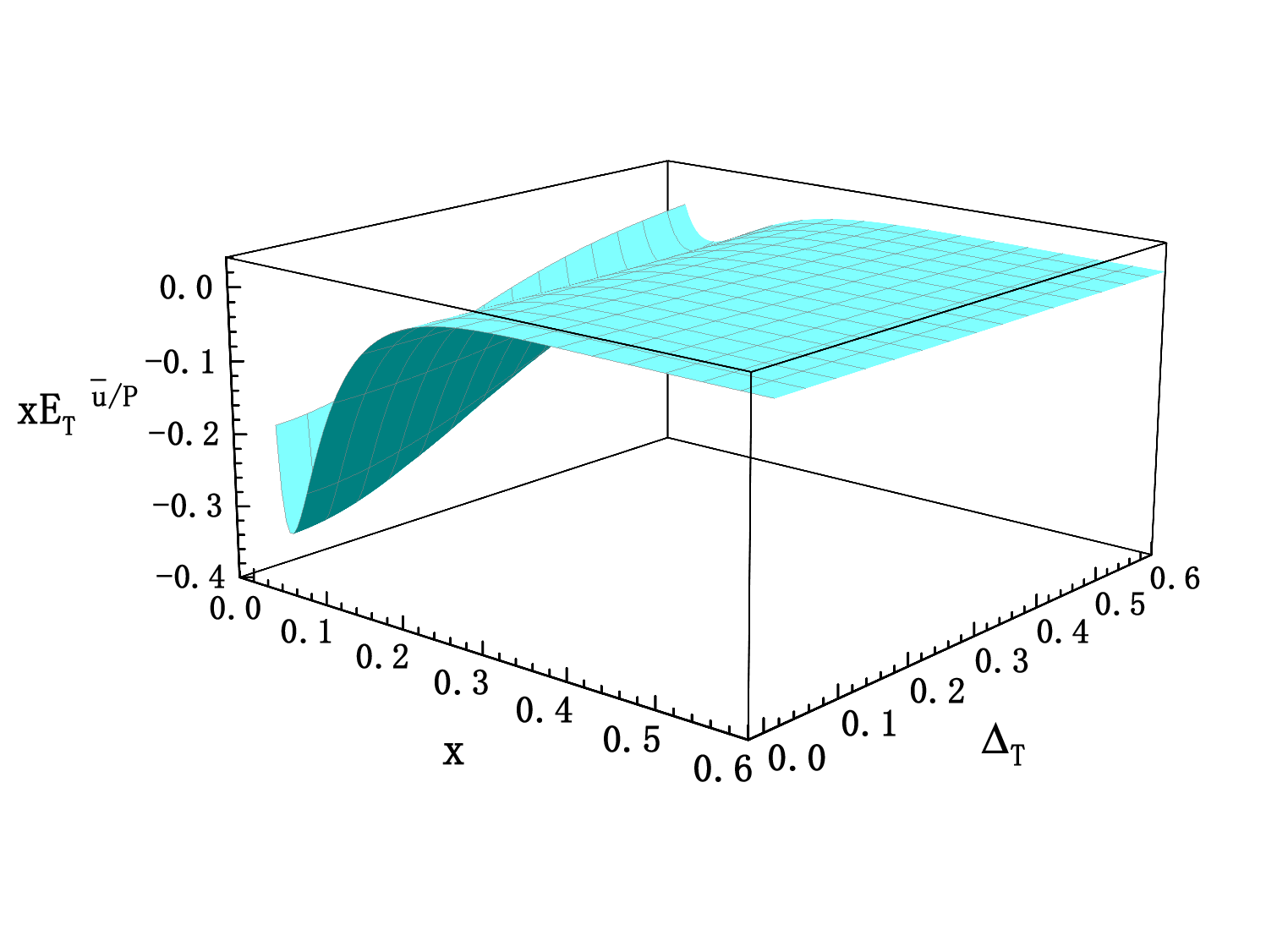}
      	\end{minipage}}
      	\subfigure{\begin{minipage}[b]{0.45\linewidth}
      			\centering
      			\includegraphics[width=\linewidth]{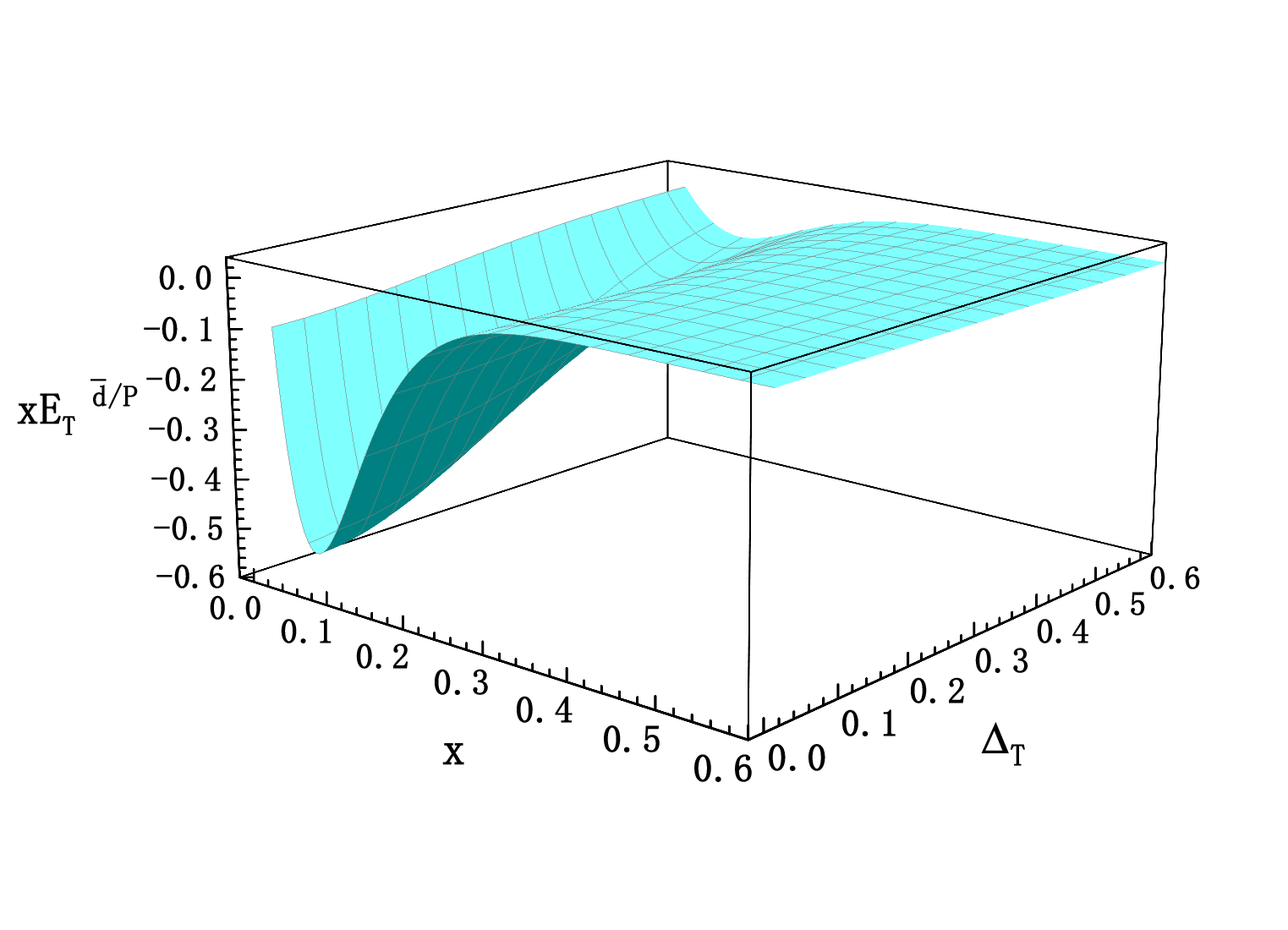}    	\end{minipage}}
      	\caption{The chiral-odd GPDs in momentum space $E^{\bar{u}/P}_T(x,0,-\boldsymbol{\Delta}_T^2)$ and $E^{\bar{d}/P}_T(x,0,-\boldsymbol{\Delta}_T^2)$ in the light-cone quark model as functions of $x$ and $\Delta_T$.} \label{Et1}      
      \end{figure*}

      \begin{figure*}[htbp]
      	\centering
      	\subfigure{\begin{minipage}[b]{0.45\linewidth}
      			\centering
      			\includegraphics[width=\linewidth]{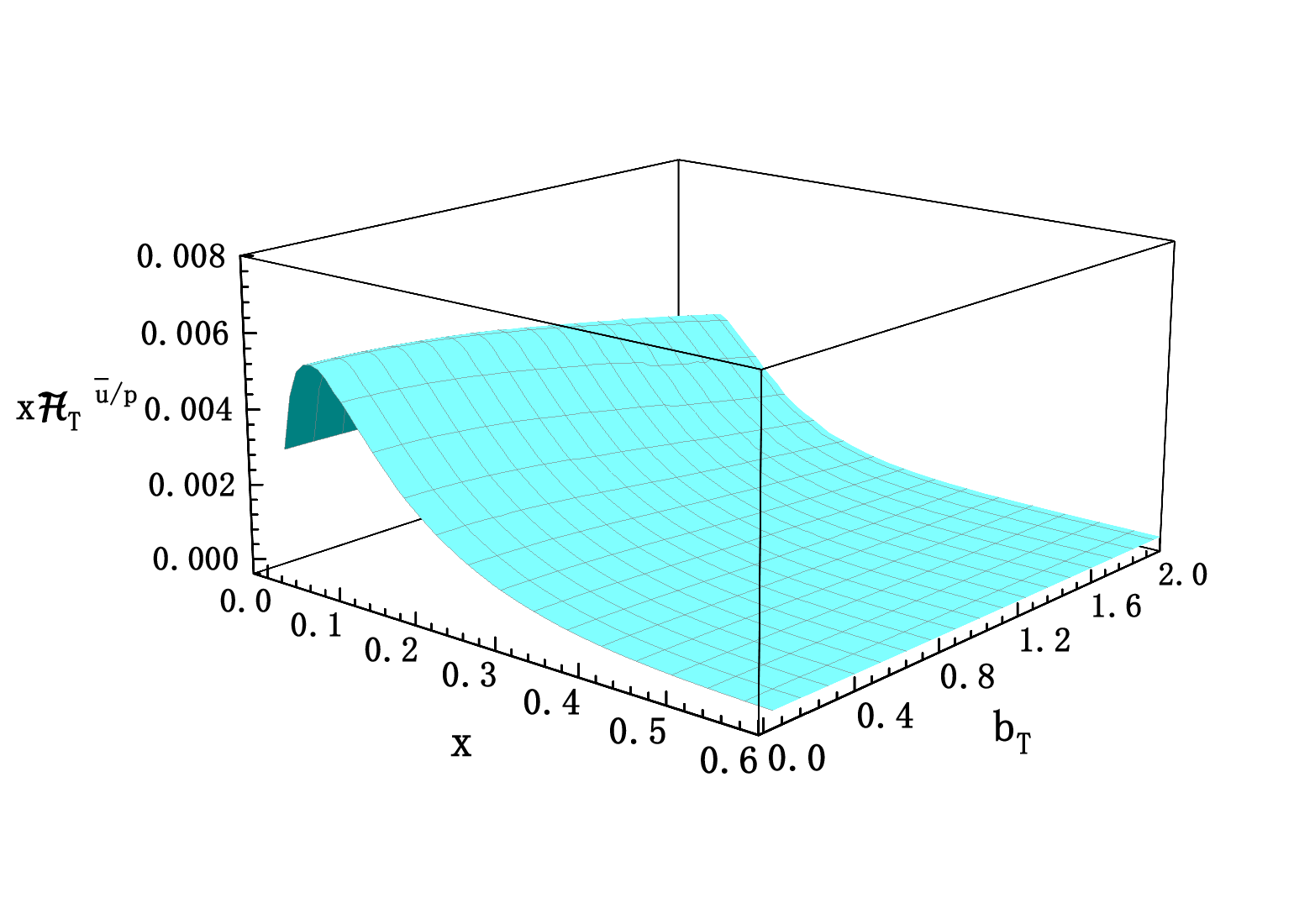}
      	\end{minipage}}
      	\subfigure{\begin{minipage}[b]{0.45\linewidth}
      			\centering
      			\includegraphics[width=\linewidth]{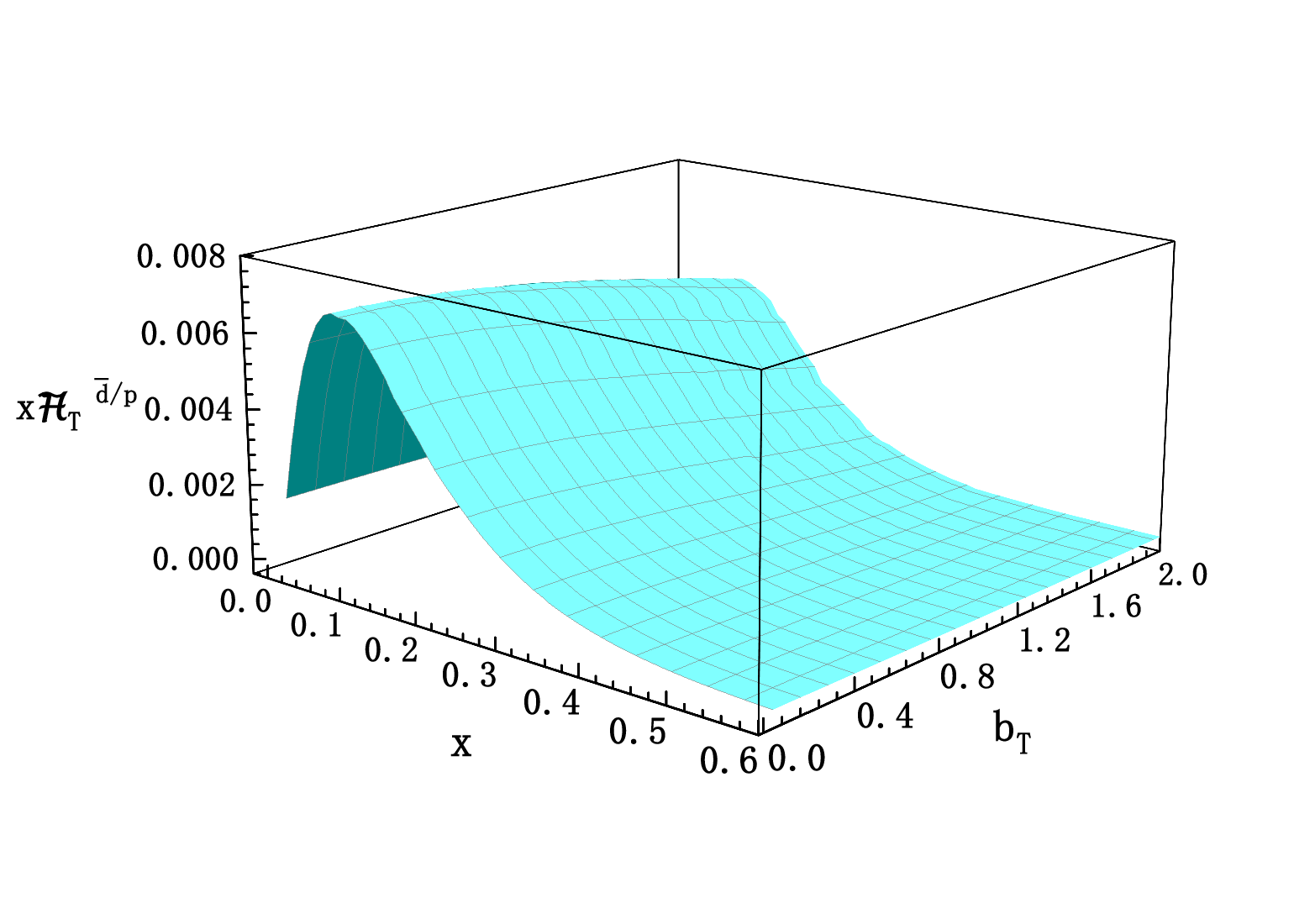}    	\end{minipage}}
      	\caption{The chiral-odd GPDs in the impact parameter space $\widetilde{\mathcal{H}}^{\bar{u}/P}_T(x, 0, \boldsymbol{b}_{T})$ and $\widetilde{\mathcal{H}}^{\bar{d}/P}_T(x, 0, \boldsymbol{b}_{T})$ in the light-cone quark model as functions of $x$ and $b_T$ .} \label{Ht3}     
      \end{figure*}
      \begin{figure*}[htbp]
      	\centering
      	\subfigure{\begin{minipage}[b]{0.45\linewidth}
      			\centering
      			\includegraphics[width=\linewidth]{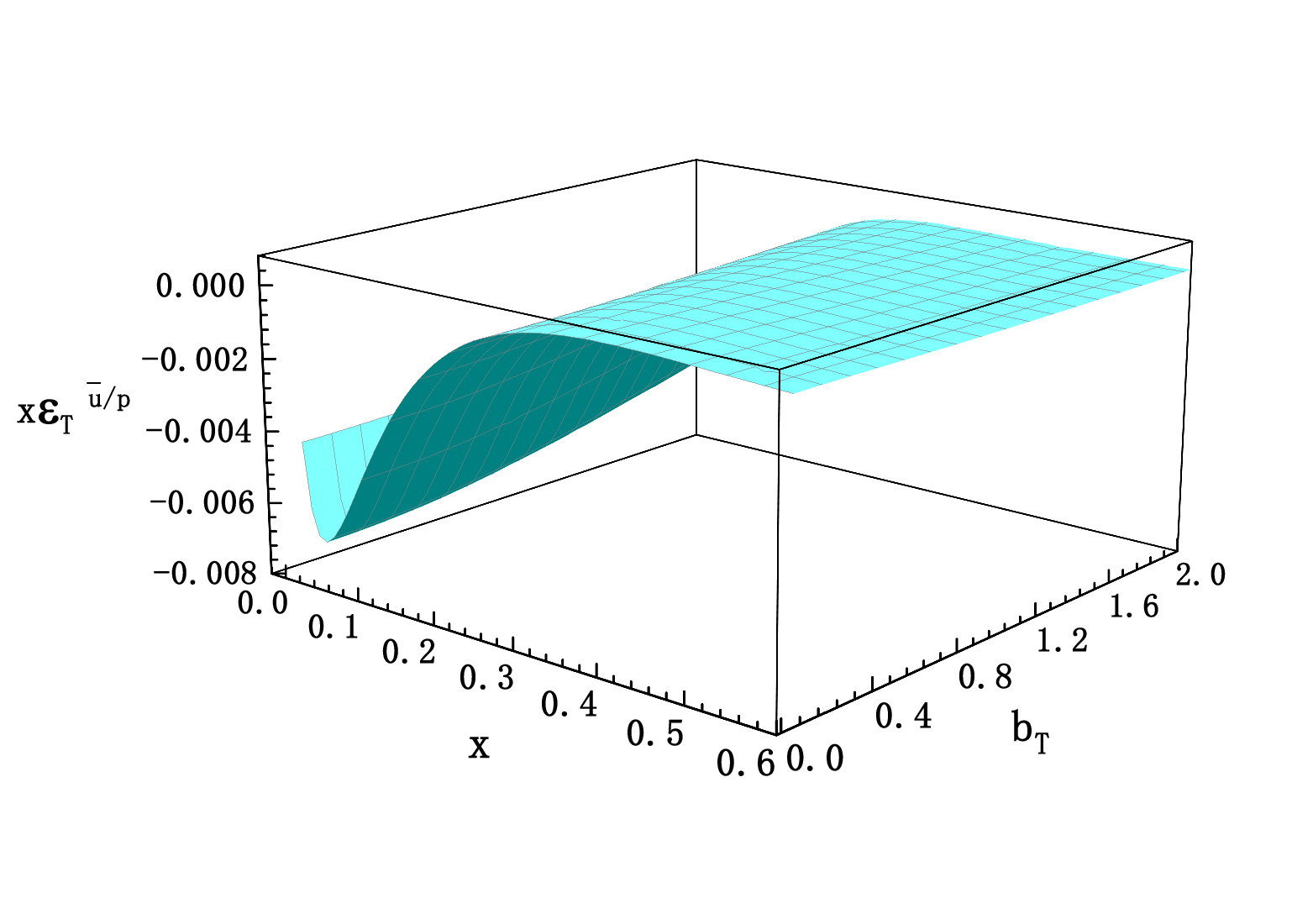}
      	\end{minipage}}
      	\subfigure{\begin{minipage}[b]{0.45\linewidth}
      			\centering
      			\includegraphics[width=\linewidth]{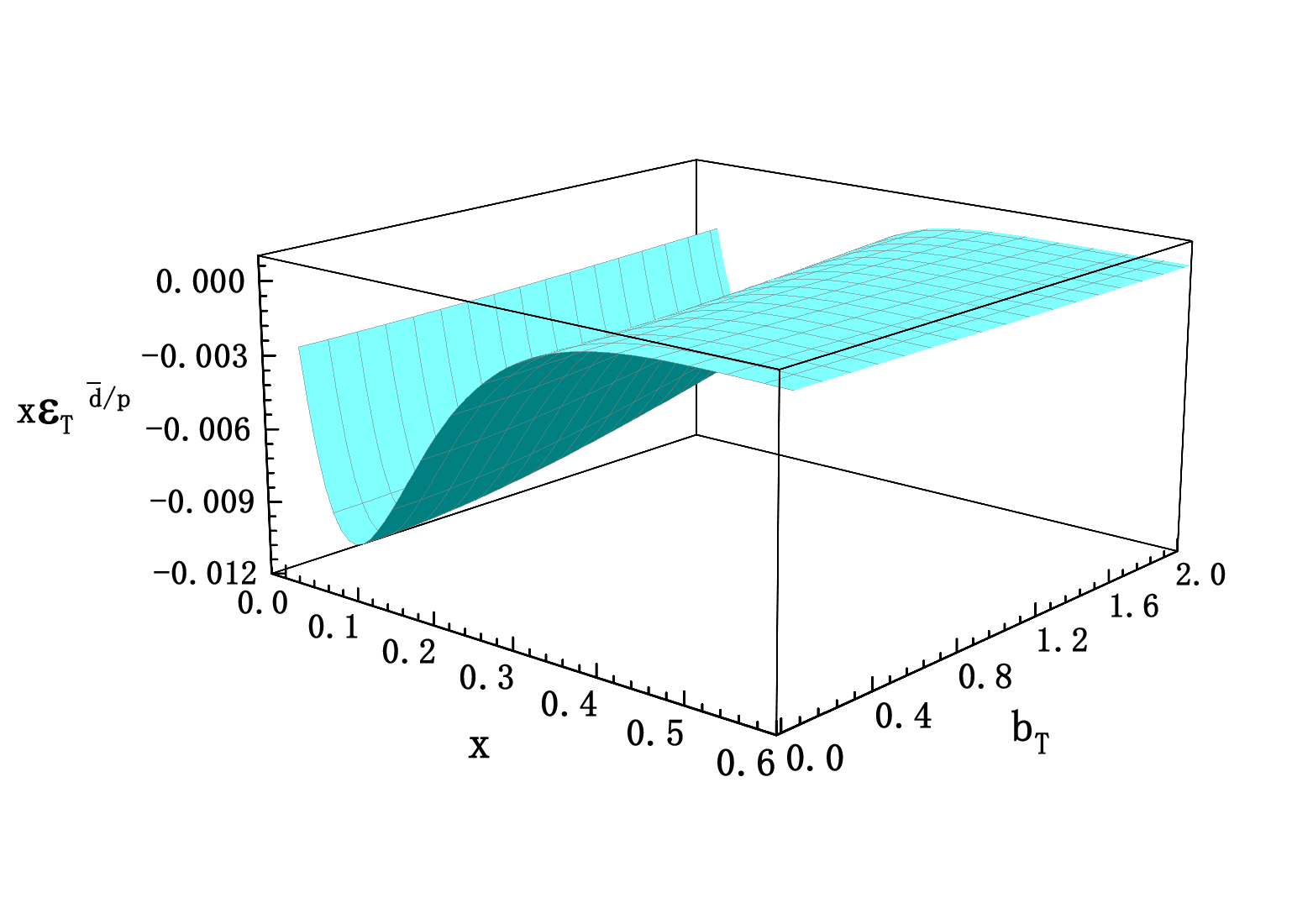}    	\end{minipage}}
      	\caption{The chiral-odd GPDs in impact parameter space $\mathcal{E}^{\bar{u}/P}_T(x, 0, \boldsymbol{b}_{T})$ and $\mathcal{E}^{\bar{d}/P}_T(x, 0, \boldsymbol{b}_{T})$  in the light-cone quark model as functions of $x$ and $b_T$ .} \label{Et3}     
    \end{figure*}

Using the values of the parameters given in Table~\ref{tab1}, we numerically calculate the sea quark chiral-odd GPDs at the model scale. In the left and right panels of Fig.~\ref{Ht1}, we plot the $\widetilde{H}_T^{\overline{q}/P}(x,0,-\boldsymbol{\Delta}_T^2)$ (multiplied with a prefactor $x$) of $\bar{u}$ and $\bar{d}$ quarks as a function of the momentum fraction $x$ and the momentum transfer $\Delta_T$, respectively. 
We find that $x\widetilde{H}_T^{\overline{q}/P}(x,0,-\boldsymbol{\Delta}_T^2)$ is sizable, The distribution peaks at around $x=0.08$, and the magnitude reaches 0.4 in maximum. In both the cases of $\bar{u}$ and $\bar{d}$, the signs of $x\widetilde{H}_T^{\overline{u}/P}$ and $x\widetilde{H}_T^{\overline{d}/P}$ are positive in the entire $x$ and $\Delta_T$ region. 
The peak of the curves is driven to the smaller $x$ region when $\Delta_T$ decreases.

In Fig.~\ref{Et1}, we plot $xE_T^{\overline{q}/P}(x,0,-\boldsymbol{\Delta}_T^2)$ of the $\bar{u}$ and $\bar{d}$ quarks as a function of $x$ and  $\Delta_T$.  
the magnitude of $xE_T^{\overline{q}/P}(x,0,-\boldsymbol{\Delta}_T^2)$ is similar to that of $x\widetilde{H}_T^{\overline{q}/P}(x,0,-\boldsymbol{\Delta}_T^2)$, but its sign is negative. 
Again, $xE_T^{\overline{u}/P}$ and $xE_T^{\overline{d}/P}$  peak at lower $x$ $(0<x<0.1)$. 
For fixed $\Delta_T$ value, $xE_T^{\overline{u}/P}$ and $xE_T^{\overline{d}/P}$ fall off monotonically with increasing values of $\Delta_T$.

In the following we consider the chiral-odd GPDs in the transverse position space. 
The GPDs in transverse position space are defined by introducing the Fourier conjugate $\boldsymbol{b}_T$ (impact parameter) of the transverse momentum transfer $\boldsymbol{\Delta}_T$ as follows~\cite{Burkardt:2002hr}:
     \begin{align}
     	\mathcal{H}_T(x, 0 , \boldsymbol{b}_{T})=\int\frac{d^{2} \boldsymbol{\Delta}_T}{(2 \pi)^{2}}   e^{-i \boldsymbol{\Delta}_T \cdot \boldsymbol{b}_T} H_{T}(x, 0, t), \\
     	\mathcal{E}_T(x, 0, \boldsymbol{b}_{T})=\int\frac{d^{2} \boldsymbol{\Delta}_T}{(2 \pi)^{2}} e^{-i \boldsymbol{\Delta}_T \cdot \boldsymbol{b}_T} E_{T}(x, 0, t), \\
     	\widetilde{\mathcal{H}}_{T}(x,0, \boldsymbol{b}_T)=\int\frac{d^{2} \boldsymbol{\Delta}_T}{(2 \pi)^{2}} e^{-i \boldsymbol{\Delta}_T \cdot \boldsymbol{b}_T} \widetilde{H}_{T}(x, 0, t).
     \end{align}
Here, $\boldsymbol{b}_T$ gives a measure of the transverse distance between the struck parton and the center of momentum of the hadron. 
In our work, we have taken $\xi=0$ which implies that the momentum transfer is completely in the transverse direction. 
In the DGLAP region $\xi<x<1$~\cite{Diehl:2003ny},  the impact parameter $b_T$ provides the transverse location of the parton where it is pulled out and put back to the nucleon, as well as the relative distance between the struck parton and the spectators.
     
In Fig.~\ref{Ht3} and ~\ref{Et3}, we present the numerical results of the chiral-odd GPDs of sea quarks in impact parameter space as functions of $x$ and $b_T$. 
We find that $x\widetilde{\mathcal{H}}^{\overline{q}/P}_T$ as well as $x\mathcal{E}^{\overline{q}/P}_T$ for $\bar{u}$ and $\bar{d}$ quarks peak at $b_T=0$. 
To be specific, for any given $x$, the peak of the curves decrease with increase of $b_T$. In addition, we find the position of the peak is located at similar $x$ region for any given $b_T$. 
In addition, $x\widetilde{\mathcal{H}}^{\overline{u}/P}_T$ and $x\widetilde{\mathcal{H}}^{\overline{d}/P}_T$ are positive while $x\mathcal{E}^{\overline{u}/P}_T$ and $x\mathcal{E}^{\overline{d}/P}_T$ are negative in the entire $x$ and $b_T$ region. For any given $x$ and $b_T$, the chiral-odd GPDs in impact parameter space of $\bar{d}$ quark is larger than that of $\bar{u}$ quark.
     
Similar to the chiral-even GPDs, the chiral-odd GPDs also have interesting interpretation in impact parameter space. 
In this space, at $\xi=0$ the chiral-odd GPDs also have a density interpretation depending on the polarization of both the active quark and the nucleon~\cite{Chakrabarti:2015ama}. 
Furthermore, certain combinations of the chiral-odd GPDs in impact parameter space affect the quark and nucleon spin correlations in different ways~\cite{Diehl:2005jf}. 
For example, the combination $H_T+\frac{\Delta_T^2}{4M^2}\widetilde{H}_T$ reduces to the transversity distribution $h_1(x)$ in the forward limit. 
The corresponding distribution in the impact parameter space $\mathcal{H}_{T}-\frac{\Delta_{b}}{4 m^{2}} \tilde{\mathcal{H}}_{T}$ represents the correlation between the transverse quark spin and the spin of the transversely polarized proton~\cite{Diehl:2005jf}, where
\begin{align}
\Delta_{b} f = \frac{\partial}{\partial b^{i}} \frac{\partial}{\partial b^{i}} f = 4 \frac{\partial}{\partial b^{2}}\left(b^{2} \frac{\partial}{\partial b^{2}}\right) f\,.
\end{align}
Similarly, $E_T + 2\tilde{H}_{T}$ describes the deformation in the center-of-momentum frame due to spin-orbit correlation. 
In the impact parameter space, $\mathcal{E}_{T}+2 \tilde{\mathcal{H}}_{T}$ describes a sideways shift in the distribution of transversely polarized quarks in an unpolarized proton. 
Besides, $E_T + 2\tilde{H}_{T}$ is related to the Boer-Mulders function and its first moment can be interpreted as the transverse anomalous magnetic moment of the proton $\kappa_T$~\cite{Burkardt:2005hp,Burkardt:2006ev}. 
Finally, the combination $\epsilon_{i j} b_{j} \frac{\partial}{\partial B}(\mathcal{E}_{T}+2 \tilde{\mathcal{H}}_{T})$ gives the spin-orbit correlation of the quarks in the proton which is a term of the spin density.
Here we write these combinations in the $\boldsymbol{b}_T$ space as~\cite{Dahiya:2007mt},
     \begin{align}
     	\notag f_T\left(x, 0 , \boldsymbol{b}_{T}\right)&=\mathcal{H}_{T}(x,0, \boldsymbol{b}_T)-\frac{\Delta_{b}}{4 M^{2}} \tilde{\mathcal{H}}_{T}(x,0, \boldsymbol{b}_T)
     	\\&=\int\frac{d^{2}\boldsymbol{\Delta}_T}{(2 \pi)^{2}}    e^{-i \boldsymbol{\Delta}_T \cdot \boldsymbol{b}_T} \left[H_{T}(x, 0, t)+\frac{\Delta_T^2}{4M^2}\widetilde{H}_T(x,0, t)\right],
    \\
 	\notag F_T\left(x, 0 , \boldsymbol{b}_{T}\right)&=\mathcal{E}_{T}(x,0, \boldsymbol{b}_T)+2 \tilde{\mathcal{H}}_{T}(x,0, \boldsymbol{b}_T)
 	\\&=\int\frac{d^{2}\boldsymbol{\Delta}_T}{(2 \pi)^{2}}    e^{-i \boldsymbol{\Delta}_T \cdot \boldsymbol{b}_T} \left[E_{T}(x, 0,t)+2 \widetilde{H}_{T}(x, 0,t)\right],
 \end{align}
     and the spin-orbit correlation
      \begin{align}
     	\notag F^i_T(x, 0 , \boldsymbol{b}_{T})&=-\epsilon^{i j} b_{j} \frac{\partial}{\partial B}\left[\mathcal{E}_{T}(x,0,\boldsymbol{b}_T)+2 \tilde{\mathcal{H}}_{T}(x,0, \boldsymbol{b}_T)\right]
     	\\\notag&=i\epsilon^{i j}\int \frac{d^{2} \boldsymbol{\Delta}_T}{(2 \pi)^{2}} \Delta_j e^{-i \boldsymbol{\Delta}_T \cdot \boldsymbol{b}_T} \left[E_{T}(x, 0, t)+2\widetilde{H}_T(x,0, t)\right]
     	\\&=-i \frac{\epsilon^{i j} b_{j}}{b} \int \frac{(\Delta)^{2} d \Delta}{2 \pi}\left[E_{T}(x, 0,t)+2 \widetilde{H}_{T}(x, 0,t)\right] J_{1}(b \Delta),
     \end{align}
     where 
     \begin{align}
     \frac{\partial}{\partial B}=2\frac{\partial}{\partial b^2}, b_1=b_T\cos\phi, b_2=b_T\sin\phi,
\end{align} 
and
     \begin{align}
     J_{n}(b \Delta)=\frac{1}{\pi} \int_{0}^{\pi} d \theta \cos (n \theta-b \Delta \sin \theta).
    \end{align}

 \begin{figure*}[t]
    	\centering
    	\subfigure{\begin{minipage}[b]{0.45\linewidth}
    			\centering
    			\includegraphics[width=\linewidth]{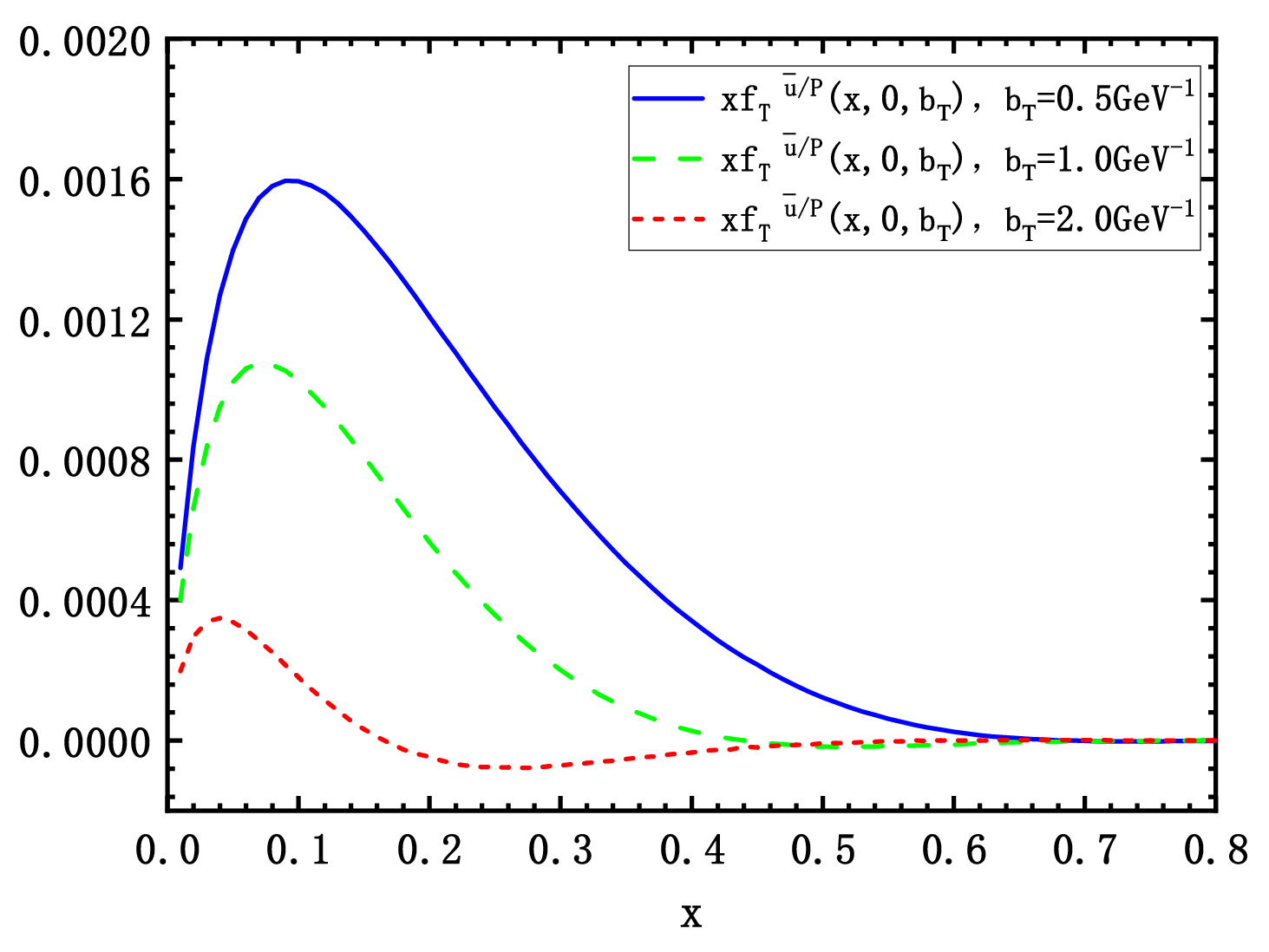}
    	\end{minipage}}
    	\subfigure{\begin{minipage}[b]{0.45\linewidth}
    			\centering
    			\includegraphics[width=\linewidth]{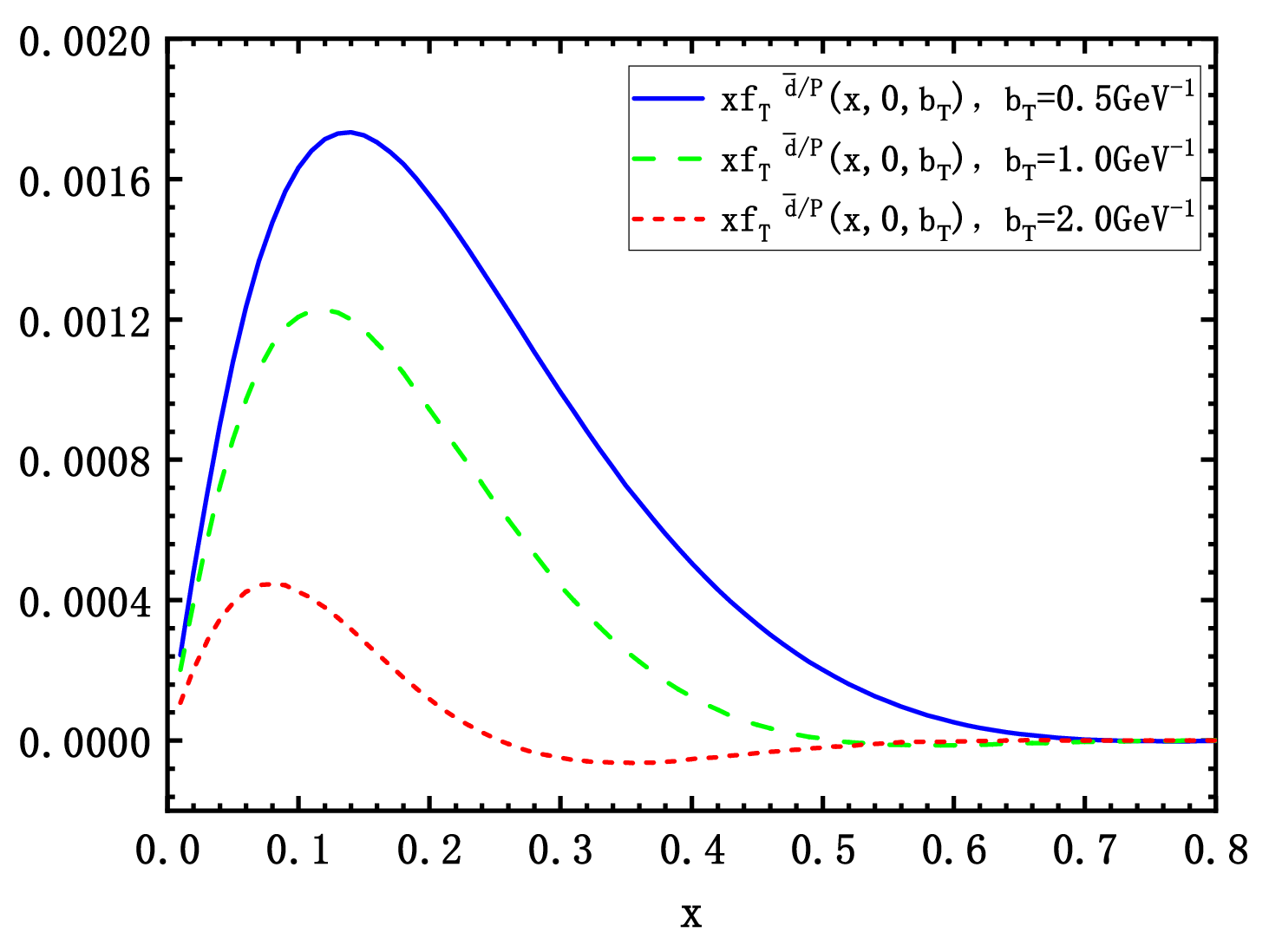}    	\end{minipage}}
    	\caption{$xf_T^{\bar{u}/P}(x,0,\boldsymbol{b}_{T})$ and $xf_T^{\bar{d}/P}(x,0,\boldsymbol{b}_{T})$ in the light-cone quark model as functions of $x$ at $b_T=0.5, 1.0$, and $2.0$ GeV$^{-1}$.} \label{ft}     
    \end{figure*}
    
    \begin{figure*}[htbp]
    	\centering
    	\subfigure{\begin{minipage}[b]{0.45\linewidth}
    			\centering
    			\includegraphics[width=\linewidth]{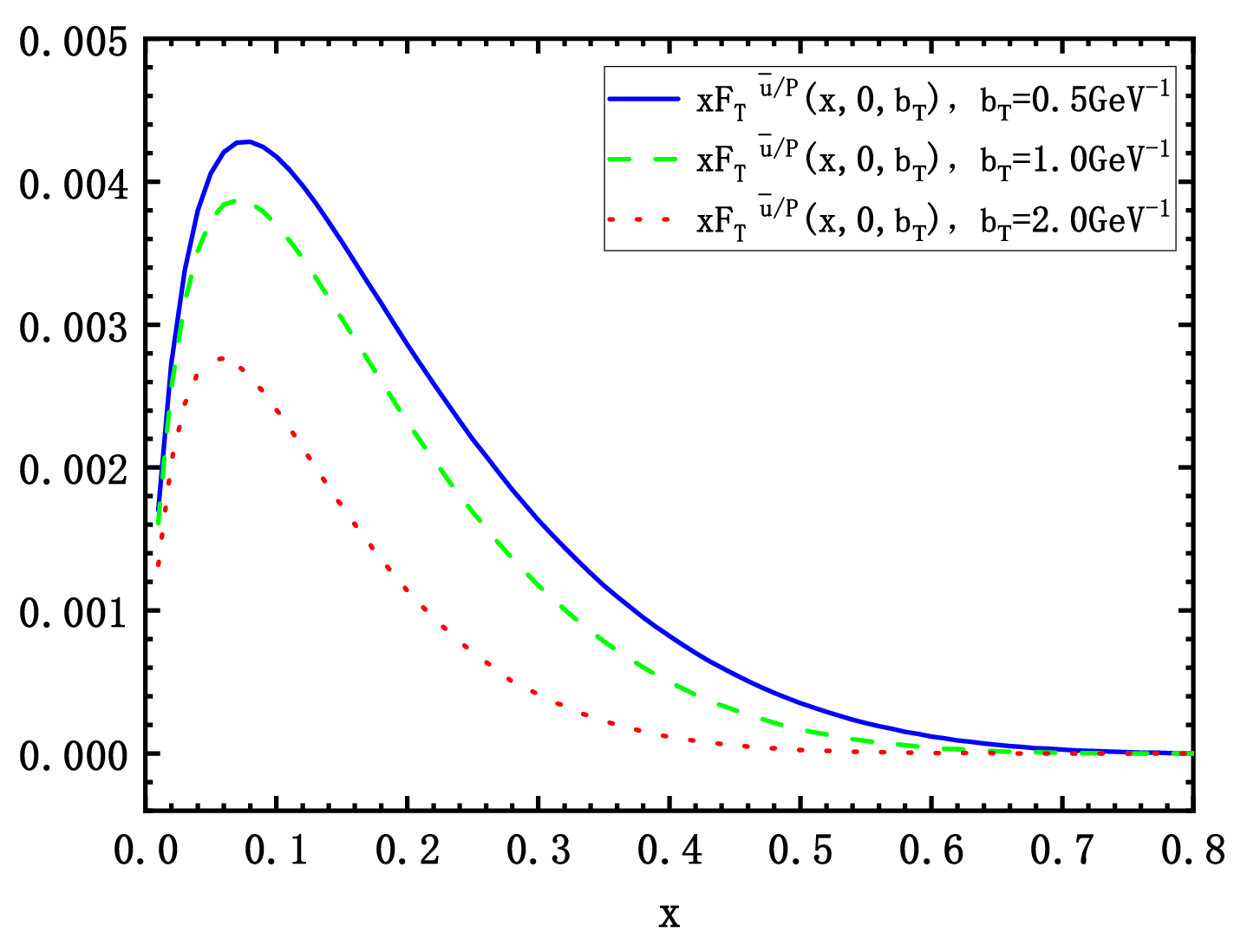}
    	\end{minipage}}
    	\subfigure{\begin{minipage}[b]{0.45\linewidth}
    			\centering
    			\includegraphics[width=\linewidth]{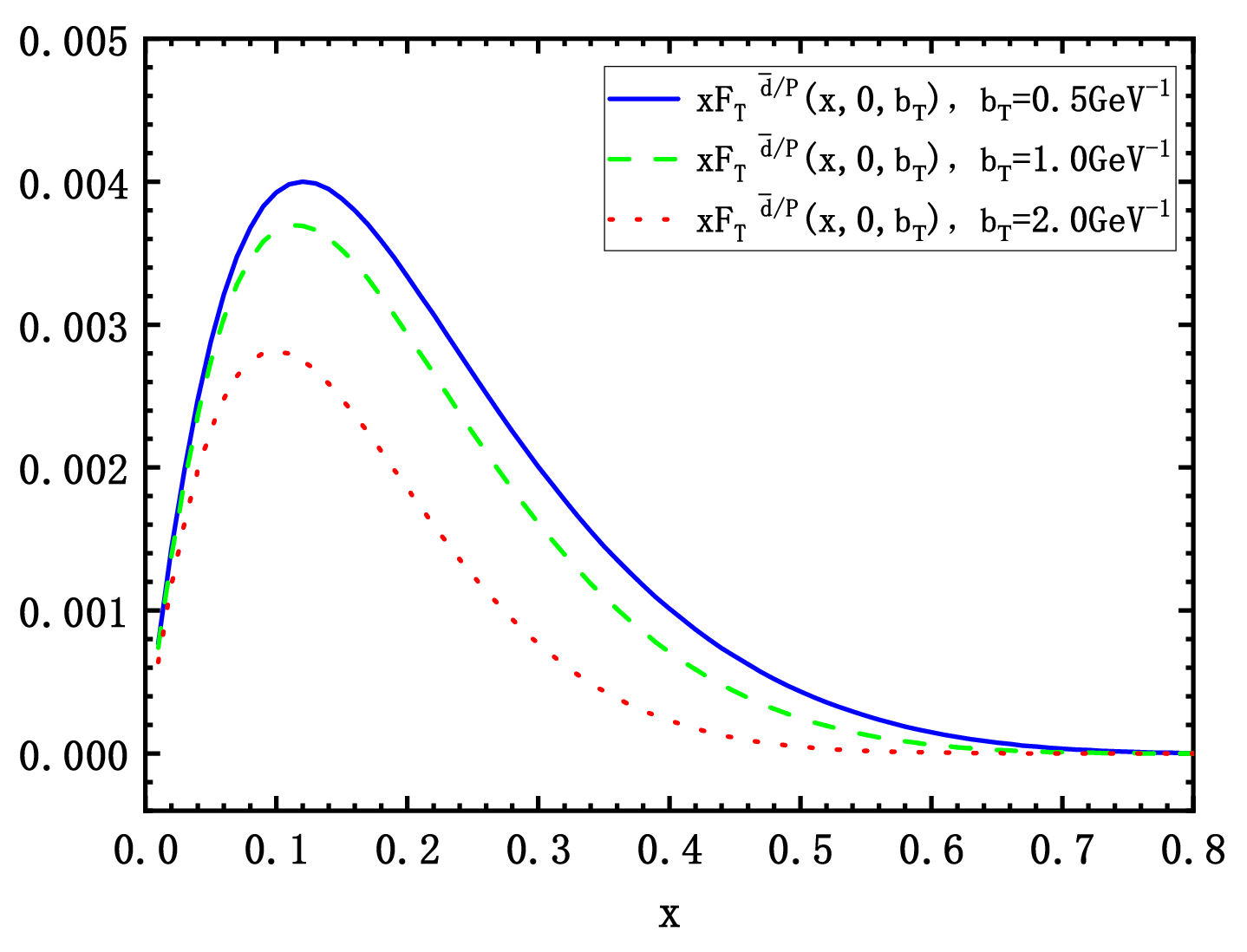}    	\end{minipage}}
    	\caption{$xF_T^{\bar{u}/P}(x,0,\boldsymbol{b}_{T})$ and $xF_T^{\bar{d}/P}(x,0,\boldsymbol{b}_{T})$  in the light-cone quark model as functions of $x$ at $b_T=0.5, 1.0$, and $2.0$ GeV$^{-1}$.} \label{Ft}     
    \end{figure*}
    
    \begin{figure*}[htbp]
    	\centering
    	\subfigure{\begin{minipage}[b]{0.45\linewidth}
    			\centering
    			\includegraphics[width=\linewidth]{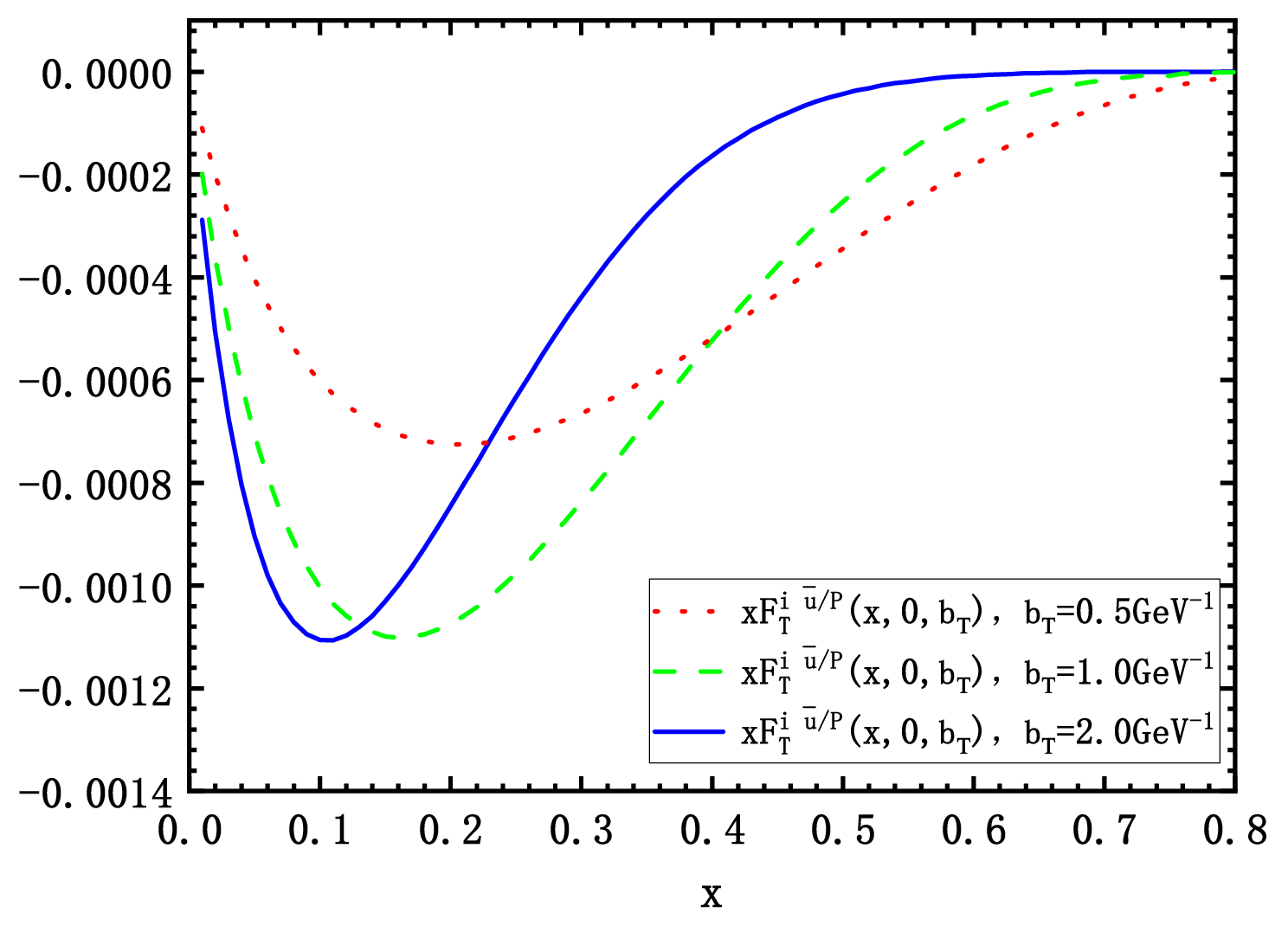}
    	\end{minipage}}
    	\subfigure{\begin{minipage}[b]{0.45\linewidth}
    			\centering
    			\includegraphics[width=\linewidth]{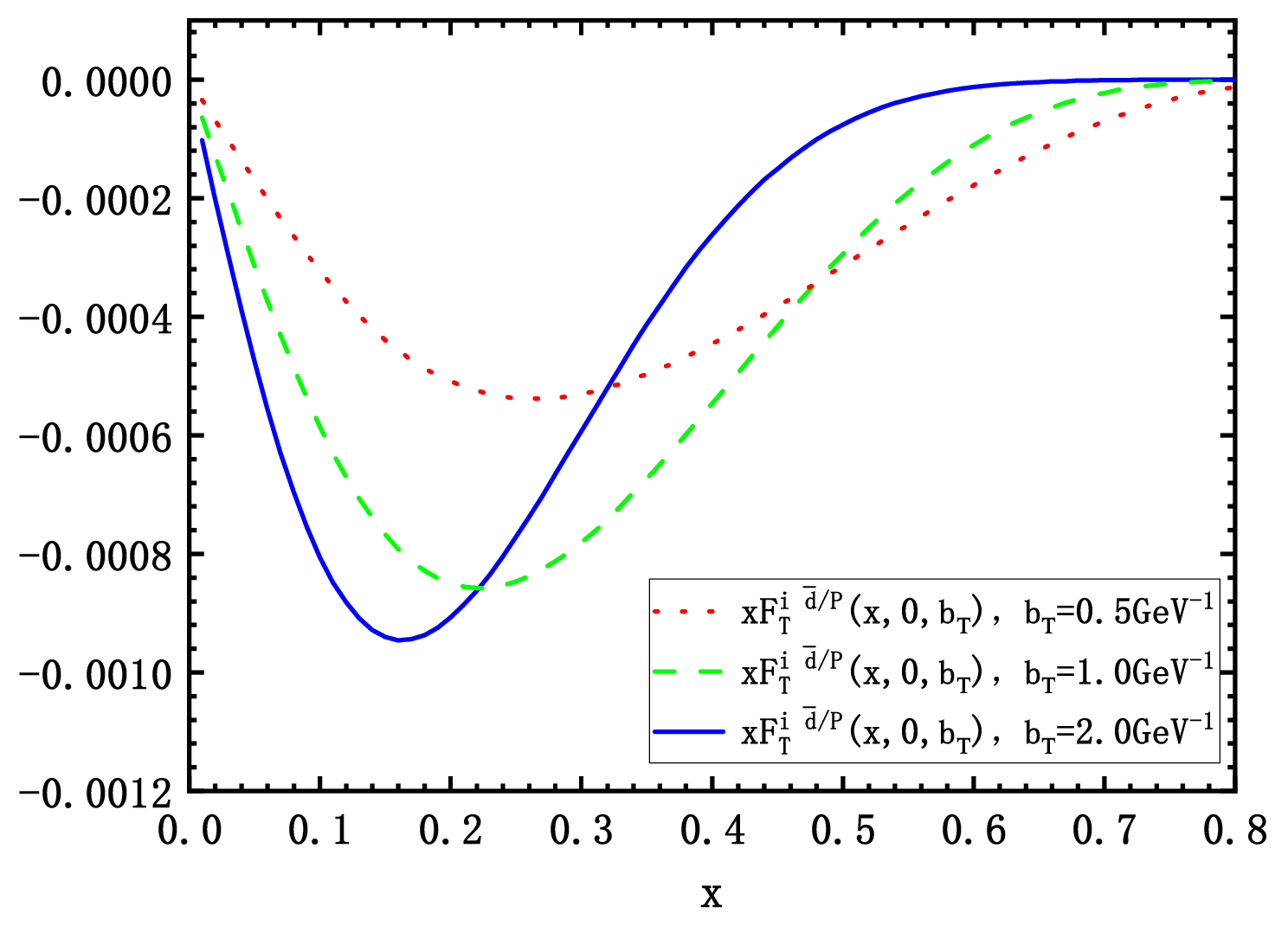}    	\end{minipage}}
    	\caption{$xF_T^{i\bar{u}/P}(x,0,\boldsymbol{b}_{T})$ and $xF_T^{i\bar{d}/P}(x,0,\boldsymbol{b}_{T})$ in the light-cone quark model as functions of $x$ at $b_T=0.5, 1.0$, and $2.0$ GeV$^{-1}$.} \label{Fti-bt}     
    \end{figure*}
    
    \begin{figure*}[htbp]
    	\centering
    	\subfigure{\begin{minipage}[b]{0.45\linewidth}
    			\centering
    			\includegraphics[width=\linewidth]{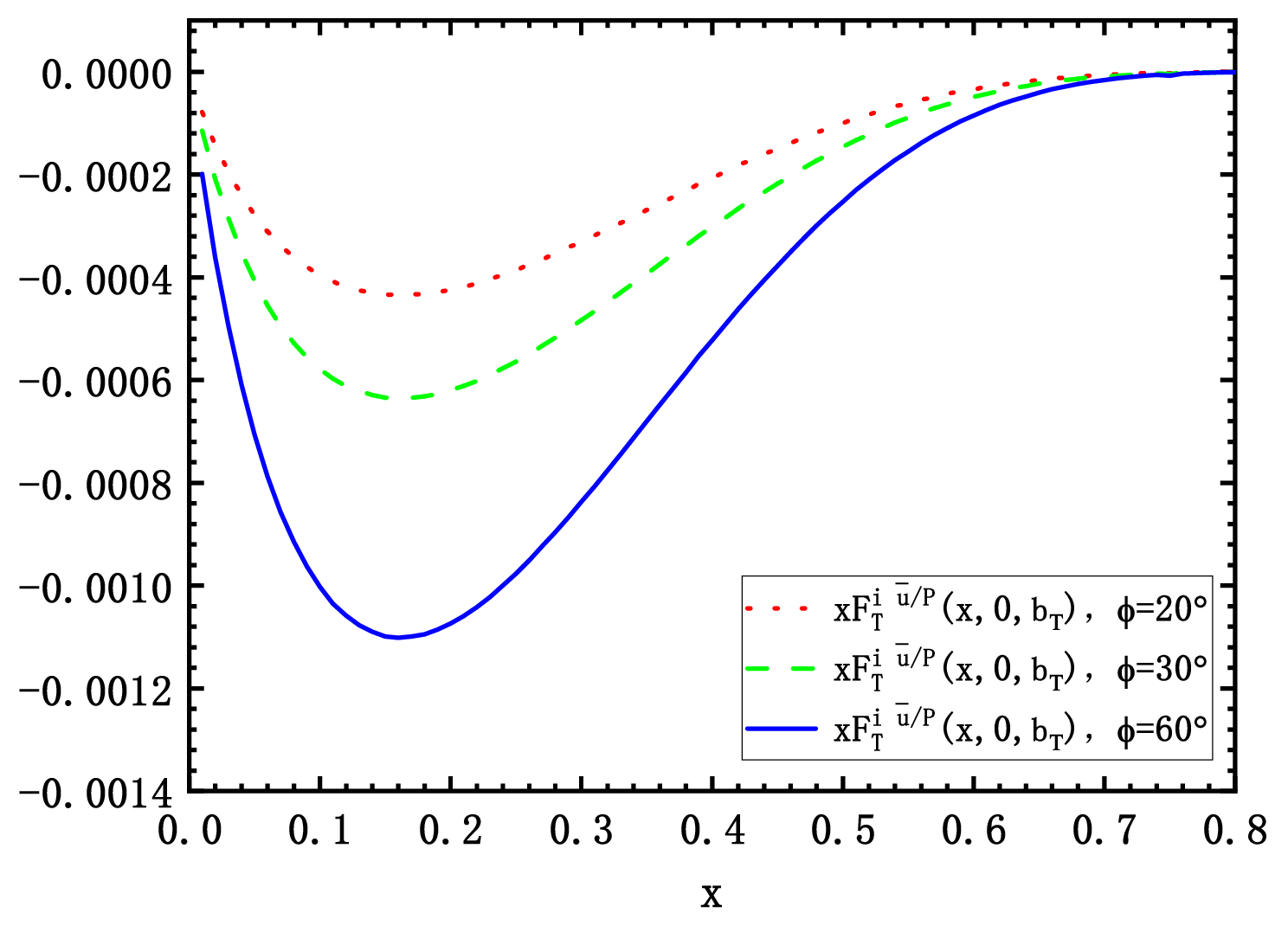}
    	\end{minipage}}
    	\subfigure{\begin{minipage}[b]{0.45\linewidth}
    			\centering
    			\includegraphics[width=\linewidth]{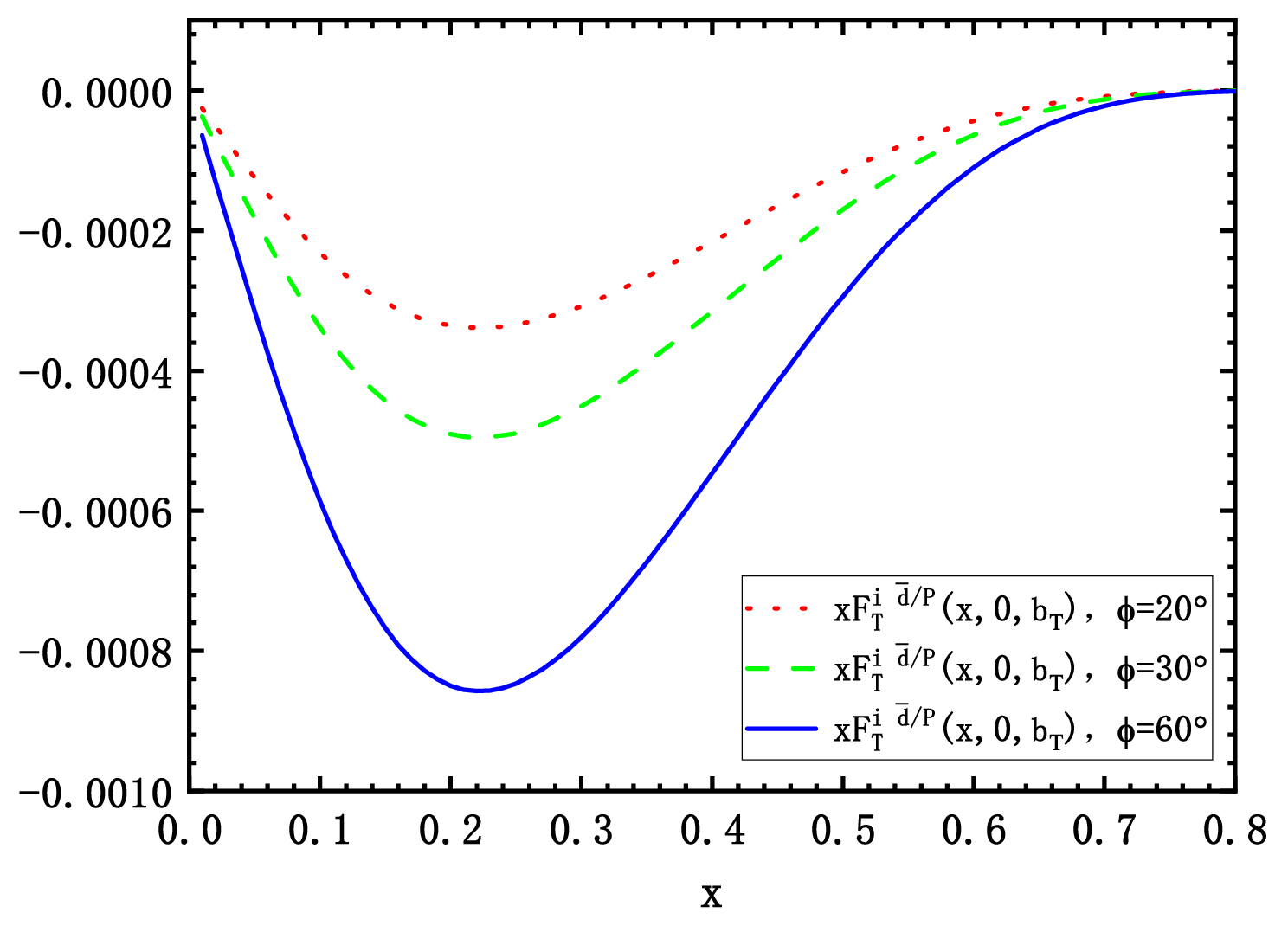}    	\end{minipage}}
    	\caption{ $xF_T^{i\bar{u}/P}(x,0,\boldsymbol{b}_{T})$ and $xF_T^{i\bar{d}/P}(x,0,\boldsymbol{b}_{T})$  in the light-cone quark model as functions of $x$ at $\phi=20^\circ, 30^\circ$ and $60^\circ$.} \label{Fti-phi}     
    \end{figure*}

In Fig.~\ref{ft}, we plot the $x$-dependence of $f_T(x, 0, \boldsymbol{b}_{T})$ of the $\bar{u}$ (left figure) and $\bar{d}$ (right figure) quarks at fixed impact parameter $b_T=0.5$ GeV$^{-1}$, $1.0$ GeV$^{-1}$ and $2.0$ GeV$^{-1}$, respectively. 
We find that $xf_T(x, 0, \boldsymbol{b}_{T})$ of the $\bar{u}$ and $\bar{d}$ quarks tend to be positive, and the large contribution is concentrated in the region $x<0.4$.
As $b_T$ increases, the size of $xf_T(x, 0, \boldsymbol{b}_{T})$ decreases and the peak of the curve is driven to the smaller $x$ region.
In Fig.~\ref{Ft}, we depict $xF_T(x, 0 , \boldsymbol{b}_{T})$ as a function of $x$ at different values of $b_T$. 
The results show that $xF_T^{\bar{u}/P}$ and $xF_T^{\bar{d}/P}$ are positive and their size decrease with increasing  $b_T$, similar to the case of $xf_T(x, 0, \boldsymbol{b}_{T})$.
Another observation is that the sizes and shapes of $xF_T^{\bar{u}/P}$ and $xF_T^{\bar{d}/P}$ are quite similar.
In Fig.~\ref{Fti-bt}, we plot $xF^i_T(x, 0 , \boldsymbol{b}_{T})$ as a function of $x$ at different values of $b_T$. Here we neglect the constant phase factor $(i)$ and take $\epsilon^{i j}=\epsilon^{1 2}$. The distribution is negative in sign for both the $\bar{u}$ and $\bar{d}$ quarks. 
As the term of $\epsilon_{i j} b_{j} \frac{\partial}{\partial B}(\mathcal{E}_{T}+2 \tilde{\mathcal{H}}_{T})$ describes the correlation between the quark spin and angular momenta, this shift clearly shows the interplay between the spin and the orbital angular momentum. 
In Fig.~\ref{Fti-phi}, we plot $xF^i_T(x, 0 , \boldsymbol{b}_{T})$ as a function of $x$ at $\phi=20^\circ$, $30^\circ$ and $60^\circ$. 
The value of $b_T$ is fixed at 1.0 GeV$^{-1}$. 
Among $f_T$, $F_T$ and $F_T^i$, the size of $F_T$ can reach 0.004 and is the largest. The size of $F_T^i$ is much smaller than that of $F_T$.
    
\section{CONCLUSION}\label{Sec4} 
     
In this work, we studied the chiral-odd GPDs of the sea quarks inside the proton using a light-cone quark model. 
We also applied the overlap representation to provide the expressions of the chiral-odd GPDs in term of the LCWF of the proton. 
To generate the sea quark degree of freedom, we treated the Fock state of proton as a composite system formed by a pion meson and a baryon, where the pion meson is composed in terms of $q\overline{q}$. 
Using the overlap representation of LCWFs, we obtained the analytic results of the chiral-odd GPDs of sea quarks at $\xi=0$. 
We found that $H_T^{\bar{q}/P}(x,0,-\boldsymbol{\Delta}_T^2)$ in our model vanishes. 
By properly choosing the values of the parameters in the model, we numerically calculated $\widetilde{H}_T^{\bar{q}/P}(x,0,-\boldsymbol{\Delta}_T^2)$  and $E_T^{\bar{q}/P}(x,0,-\boldsymbol{\Delta}_T^2)$ for $\bar{q}={\bar{u}}$ and $\bar{d}$. 
We found that the two chiral-odd are sizable, and $\widetilde{H}_T^{\bar{q}/P}(x,0,-\boldsymbol{\Delta}_T^2)$ is positive while $E_T^{\bar{q}/P}(x,0,-\boldsymbol{\Delta}_T^2)$ is negative. 
We also calculated $\mathcal{H}^{\bar{q}/P}_{T}(x,0, \boldsymbol{b}_T)$, $\mathcal{E}^{\bar{q}/P}_{T}(x,0, \boldsymbol{b}_T)$ and $\widetilde{\mathcal{H}}^{\bar{q}/P}_{T}(x,0, \boldsymbol{b}_T)$ which are the distributions in the impact parameter space. 
The numerical results show that these distributions decrease with increasing $b_T$.
To study the spin-orbit correlation effect of sea quarks quantitatively, we estimated the combinations $\mathcal{H}_{T}-\frac{\Delta_{b}}{4 M^{2}} \tilde{\mathcal{H}}_{T}$, $\mathcal{E}_{T}+2 \tilde{\mathcal{H}}_{T}$ and $\epsilon_{i j} b_{j} \frac{\partial}{\partial B}(\mathcal{E}_{T}+2 \tilde{\mathcal{H}}_{T})$. 
Among them, the combination $\mathcal{E}_{T}+2 \tilde{\mathcal{H}}_{T}$ representing the sideways shift of the transversely polarized quarks in an unpolarized proton has a size of 0.004 in maximum, showing that the spin-orbital correlation of the sea quarks may not be neglected.
Our study may provide useful information about the sea quarks inside the proton in the transverse-momentum space as well as the impact-parameter space.


\begin{thebibliography}{99}

		\bibitem{Muller:1994ses}
		D.~M\"uller, D.~Robaschik, B.~Geyer, F.~M.~Dittes and J.~Ho\v{r}ej\v{s}i,
		Fortsch. Phys. \textbf{42}, 101-141 (1994)
		[arXiv:hep-ph/9812448 [hep-ph]].

		\bibitem{Ji:1996nm}
		X.~D.~Ji,
		Phys. Rev. D \textbf{55} (1997), 7114-7125
		[arXiv:hep-ph/9609381 [hep-ph]].

		\bibitem{Radyushkin:1997ki}
		A.~V.~Radyushkin,
		Phys. Rev. D \textbf{56} (1997), 5524-5557
		[arXiv:hep-ph/9704207 [hep-ph]].

		\bibitem{Diehl:2015uka}
		M.~Diehl,
		Eur. Phys. J. A \textbf{52} (2016) no.6, 149
		[arXiv:1512.01328 [hep-ph]].

		\bibitem{Radyushkin:1996nd}
		A.~V.~Radyushkin,
		Phys. Lett. B \textbf{380} (1996), 417-425
		[arXiv:hep-ph/9604317 [hep-ph]].

		\bibitem{Belitsky:2001ns}
		A.~V.~Belitsky, D.~Mueller and A.~Kirchner,
		Nucl. Phys. B \textbf{629} (2002), 323-392
		[arXiv:hep-ph/0112108 [hep-ph]].

		\bibitem{Goloskokov:2006hr}
		S.~V.~Goloskokov and P.~Kroll,
		Eur. Phys. J. C \textbf{50} (2007), 829-842
		[arXiv:hep-ph/0611290 [hep-ph]].

		\bibitem{Goloskokov:2007nt}
		S.~V.~Goloskokov and P.~Kroll,
		Eur. Phys. J. C \textbf{53} (2008), 367-384
		[arXiv:0708.3569 [hep-ph]].

		\bibitem{Goloskokov:2009ia}
		S.~V.~Goloskokov and P.~Kroll,
		Eur. Phys. J. C \textbf{65} (2010), 137-151
		[arXiv:0906.0460 [hep-ph]].

		\bibitem{Goloskokov:2011rd}
		S.~V.~Goloskokov and P.~Kroll,
		Eur. Phys. J. A \textbf{47} (2011), 112
		[arXiv:1106.4897 [hep-ph]].

		\bibitem{Sehgal:1974rz}
		L.~M.~Sehgal,
		Phys. Rev. D \textbf{10} (1974), 1663
		[erratum: Phys. Rev. D \textbf{11} (1975), 2016]

		\bibitem{Kroll:2020jat}
		P.~Kroll,
		Mod. Phys. Lett. A \textbf{35}, no.12, 2050093 (2020)
		[arXiv:2001.01919 [hep-ph]].

		\bibitem{Brodsky:2000ii}
		S.~J.~Brodsky, D.~S.~Hwang, B.~Q.~Ma and I.~Schmidt,
		Nucl. Phys. B \textbf{593} (2001), 311-335
		[arXiv:hep-th/0003082 [hep-th]].

		\bibitem{Kumar:2014osa}
		N.~Kumar and H.~Dahiya,
		Mod. Phys. Lett. A \textbf{29} (2014), 1450118
		[arXiv:1405.7176 [hep-ph]].

		\bibitem{Miller:2010nz}
		G.~A.~Miller,
		Ann. Rev. Nucl. Part. Sci. \textbf{60} (2010), 1-25
		[arXiv:1002.0355 [nucl-th]].

		\bibitem{Kumar:2014coa}
		N.~Kumar and H.~Dahiya,
		Phys. Rev. D \textbf{90} (2014) no.9, 094030
		[arXiv:1411.0817 [hep-ph]].
		
		\bibitem{Diehl:2005jf}
		M.~Diehl and P.~Hagler,
		Eur. Phys. J. C \textbf{44} (2005), 87-101
		[arXiv:hep-ph/0504175 [hep-ph]].

		\bibitem{Burkardt:2005hp}
		M.~Burkardt,
		Phys. Rev. D \textbf{72} (2005), 094020
		[arXiv:hep-ph/0505189 [hep-ph]].
		

	    \bibitem{Pisarski:2000eq}
	    R.~D.~Pisarski,
	    Phys. Rev. D \textbf{62} (2000), 111501
	    [arXiv:hep-ph/0006205 [hep-ph]].

	   \bibitem{Burkardt:2002hr}
	   M.~Burkardt,
	   Int. J. Mod. Phys. A \textbf{18} (2003), 173-208
	   [arXiv:hep-ph/0207047 [hep-ph]].

	    \bibitem{H1:1999pji}
	    C.~Adloff \textit{et al.} [H1],
	    Eur. Phys. J. C \textbf{13} (2000), 371-396
	    [arXiv:hep-ex/9902019 [hep-ex]].

	   \bibitem{H1:2001nez}
	   C.~Adloff \textit{et al.} [H1],
	   Phys. Lett. B \textbf{517} (2001), 47-58
	   [arXiv:hep-ex/0107005 [hep-ex]].

	    \bibitem{H1:2005gdw}
	    A.~Aktas \textit{et al.} [H1],
	    Eur. Phys. J. C \textbf{44} (2005), 1-11
	    [arXiv:hep-ex/0505061 [hep-ex]].

	   \bibitem{ZEUS:1998xpo}
	   J.~Breitweg \textit{et al.} [ZEUS],
	   Eur. Phys. J. C \textbf{6} (1999), 603-627
	   [arXiv:hep-ex/9808020 [hep-ex]].

	   \bibitem{ZEUS:2003pwh}
	   S.~Chekanov \textit{et al.} [ZEUS],
	   Phys. Lett. B \textbf{573} (2003), 46-62
	   [arXiv:hep-ex/0305028 [hep-ex]].

	   \bibitem{HERMES:2001bob}
	   A.~Airapetian \textit{et al.} [HERMES],
	   Phys. Rev. Lett. \textbf{87} (2001), 182001
	   [arXiv:hep-ex/0106068 [hep-ex]].

	   \bibitem{HERMES:2011bou}
	   A.~Airapetian \textit{et al.} [HERMES],
	   Phys. Lett. B \textbf{704} (2011), 15-23
	   [arXiv:1106.2990 [hep-ex]].

	   \bibitem{HERMES:2012gbh}
	   A.~Airapetian \textit{et al.} [HERMES],
	   JHEP \textbf{07} (2012), 032
	   [arXiv:1203.6287 [hep-ex]].

	   \bibitem{dHose:2004usi}
	   N.~d'Hose, E.~Burtin, P.~A.~M.~Guichon and J.~Marroncle,
	   Eur. Phys. J. A \textbf{19S1} (2004), 47-53.

	   \bibitem{CLAS:2001wjj}
	   S.~Stepanyan \textit{et al.} [CLAS],
	   Phys. Rev. Lett. \textbf{87} (2001), 182002
	   [arXiv:hep-ex/0107043 [hep-ex]].

	    \bibitem{Polyakov:1998ze}
	   M.~V.~Polyakov,
	   Nucl. Phys. B \textbf{555} (1999), 231
	   [arXiv:hep-ph/9809483 [hep-ph]].

	   \bibitem{Collins:1996fb}
	   J.~C.~Collins, L.~Frankfurt and M.~Strikman,
	   Phys. Rev. D \textbf{56} (1997), 2982-3006
	   [arXiv:hep-ph/9611433 [hep-ph]].

	   \bibitem{Ahmad:2008hp}
	   S.~Ahmad, G.~R.~Goldstein and S.~Liuti,
	   Phys. Rev. D \textbf{79} (2009), 054014
	   [arXiv:0805.3568 [hep-ph]].

	   \bibitem{Goldstein:2010gu}
	   G.~R.~Goldstein, J.~O.~Hernandez and S.~Liuti,
	   Phys. Rev. D \textbf{84} (2011), 034007
	   [arXiv:1012.3776 [hep-ph]].

	   \bibitem{Boussarie:2016qop}
	   R.~Boussarie, B.~Pire, L.~Szymanowski and S.~Wallon,
	   JHEP \textbf{02} (2017), 054
	   [erratum: JHEP \textbf{10} (2018), 029]
	   [arXiv:1609.03830 [hep-ph]].

	    \bibitem{Ivanov:2002jj}
	   D.~Y.~Ivanov, B.~Pire, L.~Szymanowski and O.~V.~Teryaev,
	   Phys. Lett. B \textbf{550} (2002), 65-76
	   [arXiv:hep-ph/0209300 [hep-ph]].

	   \bibitem{Enberg:2006he}
	   R.~Enberg, B.~Pire and L.~Szymanowski,
	   Eur. Phys. J. C \textbf{47} (2006), 87-94
	   [arXiv:hep-ph/0601138 [hep-ph]].

	   \bibitem{Cosyn:2020kfe}
	   W.~Cosyn, B.~Pire and L.~Szymanowski,
	   Phys. Rev. D \textbf{102} (2020) no.5, 054003
	   [arXiv:2007.01923 [hep-ph]].

	   \bibitem{COMPASS:2013fsk}
	   C.~Adolph \textit{et al.} [COMPASS],
	   Phys. Lett. B \textbf{731} (2014), 19-26
	   [arXiv:1310.1454 [hep-ex]].

\bibitem{Goloskokov:2013mba}
S.~V.~Goloskokov and P.~Kroll,
Eur. Phys. J. C \textbf{74}, 2725 (2014)
[arXiv:1310.1472 [hep-ph]].

	    \bibitem{Goldstein:2013gra}
	   G.~R.~Goldstein, J.~O.~Gonzalez Hernandez and S.~Liuti,
	   Phys. Rev. D \textbf{91} (2015) no.11, 114013
	   [arXiv:1311.0483 [hep-ph]].

	   \bibitem{Diehl:2007hd}
	   M.~Diehl and W.~Kugler,
	   Eur. Phys. J. C \textbf{52} (2007), 933-966
	   [arXiv:0708.1121 [hep-ph]].

	   \bibitem{Duplancic:2016bge}
	   G.~Duplan\v{c}i\'c, D.~M\"uller and K.~Passek-Kumeri\v{c}ki,
	   Phys. Lett. B \textbf{771} (2017), 603-610
	   [arXiv:1612.01937 [hep-ph]].

	   \bibitem{Siddikov:2019ahb}
	   M.~Siddikov and I.~Schmidt,
	   Phys. Rev. D \textbf{99} (2019) no.11, 116005
	   [arXiv:1904.04252 [hep-ph]].

	    \bibitem{CLAS:2014jpc}
	    I.~Bedlinskiy \textit{et al.} [CLAS],
	    Phys. Rev. C \textbf{90} (2014) no.2, 025205
	    [arXiv:1405.0988 [nucl-ex]].

	    \bibitem{CLAS:2017jjr}
	    I.~Bedlinskiy \textit{et al.} [CLAS],
	    Phys. Rev. C \textbf{95} (2017) no.3, 035202
	    [arXiv:1703.06982 [nucl-ex]].

	    \bibitem{Kim:2015pkf}
	    A.~Kim, H.~Avakian, V.~Burkert, K.~Joo, W.~Kim, K.~P.~Adhikari, Z.~Akbar, S.~Anefalos Pereira, R.~A.~Badui and M.~Battaglieri, \textit{et al.}
	    Phys. Lett. B \textbf{768} (2017), 168-173
	    [arXiv:1511.03338 [nucl-ex]].

	    \bibitem{CLAS:2019uzc}
	    B.~Zhao \textit{et al.} [CLAS],
	    Phys. Lett. B \textbf{789} (2019), 426-431.

	    \bibitem{Beiyad:2010qg}
	    M.~E.~Beiyad, B.~Pire, M.~Segond, L.~Szymanowski and S.~Wallon,
	    PoS \textbf{DIS2010} (2010), 252
	    [arXiv:1006.0740 [hep-ph]].

	    \bibitem{Duplancic:2023kwe}
	    G.~Duplan\v{c}i\'c, S.~Nabeebaccus, K.~Passek-Kumeri\v{c}ki, B.~Pire, L.~Szymanowski and S.~Wallon,
	    Phys. Rev. D \textbf{107} (2023) no.9, 094023
	    [arXiv:2302.12026 [hep-ph]].

	    \bibitem{Pire:2019hos}
	    B.~Pire, L.~Szymanowski and S.~Wallon,
	    Phys. Rev. D \textbf{101} (2020) no.7, 074005
	    [erratum: Phys. Rev. D \textbf{103} (2021) no.5, 059901]
	    [arXiv:1912.10353 [hep-ph]].

	    \bibitem{Scopetta:2005fg}
	    S.~Scopetta,
	    Phys. Rev. D \textbf{72} (2005), 117502
	    [arXiv:hep-ph/0509287 [hep-ph]].

	    \bibitem{Pasquini:2005dk}
	    B.~Pasquini, M.~Pincetti and S.~Boffi,
	    Phys. Rev. D \textbf{72} (2005), 094029
	    [arXiv:hep-ph/0510376 [hep-ph]].

	    \bibitem{Pincetti:2006hc}
	    M.~Pincetti, B.~Pasquini and S.~Boffi,
	    Czech. J. Phys. \textbf{56} (2006), F229-F236
	    [arXiv:hep-ph/0610051 [hep-ph]].

	   \bibitem{Chakrabarti:2015ama}
	   D.~Chakrabarti and C.~Mondal,
	   Phys. Rev. D \textbf{92} (2015) no.7, 074012
	   [arXiv:1509.00598 [hep-ph]].

	    \bibitem{Chakrabarti:2008mw}
	    D.~Chakrabarti, R.~Manohar and A.~Mukherjee,
	    Phys. Rev. D \textbf{79} (2009), 034006
	    [arXiv:0811.0521 [hep-ph]].

	    \bibitem{Dahiya:2007mt}
	    H.~Dahiya and A.~Mukherjee,
	    Phys. Rev. D \textbf{77} (2008), 045032
	    [arXiv:0711.1566 [hep-ph]].

	   \bibitem{Kumar:2015yta}
	   N.~Kumar and H.~Dahiya,
	   Phys. Rev. D \textbf{91} (2015) no.11, 114031
	   [arXiv:1506.03168 [hep-ph]].

	   \bibitem{Goldstein:2014aja}
	   G.~R.~Goldstein, J.~O.~Gonzalez Hernandez and S.~Liuti,
	   [arXiv:1401.0438 [hep-ph]].

	   \bibitem{Gockeler:2005aw}
	   M.~Gockeler \textit{et al.} [QCDSF],
	   Nucl. Phys. A \textbf{755} (2005), 537-544
	   [arXiv:hep-lat/0501029 [hep-lat]].

	   \bibitem{Gockeler:2005cj}
	   M.~Gockeler \textit{et al.} [QCDSF and UKQCD],
	   Phys. Lett. B \textbf{627} (2005), 113-123
	   [arXiv:hep-lat/0507001 [hep-lat]].

	   \bibitem{QCDSF-UKQCD:2008gss}
	   D.~Brommel \textit{et al.} [QCDSF-UKQCD],
	   Prog. Part. Nucl. Phys. \textbf{61} (2008), 73-80.

	   \bibitem{Constantinou:2014fka}
	   M.~Constantinou, R.~Horsley, H.~Panagopoulos, H.~Perlt, P.~E.~L.~Rakow, G.~Schierholz, A.~Schiller and J.~M.~Zanotti,
	   Phys. Rev. D \textbf{91} (2015) no.1, 014502
	   [arXiv:1408.6047 [hep-lat]].

	   \bibitem{Lin:2017snn}
	   H.~W.~Lin, E.~R.~Nocera, F.~Olness, K.~Orginos, J.~Rojo, A.~Accardi, C.~Alexandrou, A.~Bacchetta, G.~Bozzi and J.~W.~Chen, \textit{et al.}
	   Prog. Part. Nucl. Phys. \textbf{100} (2018), 107-160
	   [arXiv:1711.07916 [hep-ph]].

	   \bibitem{Constantinou:2020hdm}
	   M.~Constantinou, A.~Courtoy, M.~A.~Ebert, M.~Engelhardt, T.~Giani, T.~Hobbs, T.~J.~Hou, A.~Kusina, K.~Kutak and J.~Liang, \textit{et al.}
	   Prog. Part. Nucl. Phys. \textbf{121} (2021), 103908
	   [arXiv:2006.08636 [hep-ph]].

      \bibitem{Brodsky:1996hc}
      S.~J.~Brodsky and B.~Q.~Ma,
      Phys. Lett. B \textbf{381}, 317-324 (1996)
      [arXiv:hep-ph/9604393 [hep-ph]].

	   \bibitem{Luan:2022fjc}
	   X.~Luan and Z.~Lu,
	   Phys. Lett. B \textbf{833} (2022), 137299
	   [arXiv:2204.06854 [hep-ph]].

	    \bibitem{Rajan:2017cpx}
	    A.~Rajan, M.~Engelhardt and S.~Liuti,
	    Phys. Rev. D \textbf{98} (2018) no.7, 074022.
	    [arXiv:1709.05770 [hep-ph]].

	   \bibitem{Diehl:2001pm}
	   M.~Diehl,
	   Eur. Phys. J. C \textbf{19} (2001), 485-492
	   [arXiv:hep-ph/0101335 [hep-ph]].

	    \bibitem{Brodsky:2000xy}
	    S.~J.~Brodsky, M.~Diehl and D.~S.~Hwang,
	    Nucl. Phys. B \textbf{596} (2001), 99-124
	    [arXiv:hep-ph/0009254 [hep-ph]].

	   \bibitem{Lepage:1980fj}
	   G.~P.~Lepage and S.~J.~Brodsky,
	   Phys. Rev. D \textbf{22} (1980), 2157.

	   \bibitem{Xiao:2003wf}
	   B.~W.~Xiao and B.~Q.~Ma,
	   Phys. Rev. D \textbf{68} (2003), 034020
	   [arXiv:hep-ph/0312162 [hep-ph]].

       \bibitem{Zeng:2023nnb}
       C.~Zeng, H.~Dong, T.~Liu, P.~Sun and Y.~Zhao,
       Phys. Rev. D \textbf{109}, no.5, 056002 (2024)
       [arXiv:2310.15532 [hep-ph]].

	   \bibitem{Gluck:1991ey}
	   M.~Gluck, E.~Reya and A.~Vogt,
	   Z. Phys. C \textbf{53} (1992), 651-656.

	    \bibitem{Martin:2009iq}
	    A.~D.~Martin, W.~J.~Stirling, R.~S.~Thorne and G.~Watt,
	    Eur. Phys. J. C \textbf{63} (2009), 189-285
	    [arXiv:0901.0002 [hep-ph]].

	    \bibitem{Diehl:2003ny}
	    M.~Diehl,
	    Phys. Rept. \textbf{388} (2003), 41-277
	    [arXiv:hep-ph/0307382 [hep-ph]].

	   \bibitem{Burkardt:2006ev}
	   M.~Burkardt,
	   Phys. Lett. B \textbf{639} (2006), 462-464.
	 \end{thebibliography}
\end{document}